\documentclass[aps,amssymb,amsmath,prd,twocolumn,
showpacs,preprintnumbers,superscriptaddress,nofootinbib,floatfix]{revtex4-1}
\usepackage{graphicx}
\usepackage{bm}
\usepackage{amssymb,amsmath}        
\usepackage{mathrsfs}
\usepackage{latexsym}
\usepackage{float}
\usepackage{color}
\usepackage[normalem]{ulem} 
\usepackage{dcolumn}
\usepackage[colorlinks=true,citecolor=blue,urlcolor=blue]{hyperref}
\usepackage[usenames,dvipsnames]{xcolor}
\usepackage[caption=false]{subfig}
\usepackage{slashed}
\usepackage[dvipsnames]{xcolor}
\usepackage{orcidlink} 
\usepackage{enumerate}
\usepackage{relsize}                  

\newcommand\eqn[1]{(\ref{#1})}      
\newcommand\Eqn[1]{Eq.~(\ref{#1})}  

\newcommand{\p}{\partial}
\newcommand{\nn}{\nonumber}
\newcommand{\be}{\begin{equation}}
\newcommand{\ee}{\end{equation}}
\newcommand{\beq}{\begin{eqnarray}}
\newcommand{\eeq}{\end{eqnarray}}

\newcommand{\nb}{n_{\rm B}}
\newcommand{\mub}{\mu_{\rm B}}
\newcommand{\nsat}{n_{\rm sat}}
\newcommand{\Msolar}{{\rm M}_{\odot}}

\newcommand{\ep}{\varepsilon}

\newcommand{\fed}{\mathcal{F}}

\newcommand{\cad}{c_{\rm ad}}

\newcommand{\eN}{e\mathsmaller{N}}
\newcommand{\bareN}{\bar{e\mathsmaller{N}}}
\newcommand{\eG}{e\mathsmaller{G}}
\newcommand{\bareG}{\bar{e\mathsmaller{G}}}
\newcommand{\eQ}{e\mathsmaller{Q}}
\newcommand{\bareQ}{\bar{e\mathsmaller{Q}}}

\newcommand{\xy}{\right|_{Y_e,Y}}
\newcommand{\ny}{\right|_{\nb,Y}}
\newcommand{\nx}{\right|_{\nb,Y_e}}
\newcommand{\xye}{\right|_{Y_e,Y^{eq}}}
\newcommand{\nye}{\right|_{\nb,Y^{eq}}}
\newcommand{\ye}{Y^{eq}}
\newcommand{\x}{\right|_{Y_e}}
\newcommand{\n}{\right|_{\nb}}

\begin{document}
\preprint{N3AS-25-003, INT-PUB-25-003}
\title{Framework for phase transitions between the Maxwell and Gibbs constructions at finite temperature} 

\author{Constantinos Constantinou
\orcidlink{0000-0002-4932-0879}}
\email{cconstantinou@ectstar.eu}
\affiliation{INFN-TIFPA, Trento Institute of Fundamental Physics and Applications, Povo, 38123 TN, Italy}
\affiliation{European Centre for Theoretical Studies in Nuclear Physics and Related Areas, Villazzano, 38123 TN, Italy}

\author{Mirco~Guerrini
\orcidlink{0009-0009-5946-3900}}
\email{mirco.guerrini@unife.it}
\affiliation{Department of Physics and Earth Science, University of Ferrara, 44122 Ferrara, Italy}
\affiliation{INFN, Sezione di Ferrara, 44122 Ferrara, Italy}

\author{Tianqi Zhao
\orcidlink{0000-0003-4704-0109}}
\email{tianqi.zhao@berkeley.edu}
\affiliation{Department of Physics and Astronomy, Ohio University,
Athens, OH~45701, USA}
\affiliation{Institute for Nuclear Theory, University of Washington, Seattle, WA~98195, USA}
\affiliation{Network for Neutrinos, Nuclear Astrophysics, and Symmetries (N3AS), University of California, Berkeley, Berkeley, CA~94720, USA}

\author{Sophia~Han
\orcidlink{0000-0002-9176-4617}}
\email{sjhan@sjtu.edu.cn}
\affiliation{Tsung-Dao Lee Institute, Shanghai Jiao Tong University, Shanghai~201210, China}
\affiliation{School of Physics and Astronomy, Shanghai Jiao Tong University, Shanghai~200240, China}
\affiliation{Network for Neutrinos, Nuclear Astrophysics, and Symmetries (N3AS), University of California, Berkeley, Berkeley, CA~94720, USA}
\affiliation{State Key Laboratory of Dark Matter Physics, Shanghai Jiao Tong University, Shanghai 201210, China}

\author{Madappa~Prakash
\orcidlink{0000-0002-9019-5029}}
\email{prakash@ohio.edu}
\affiliation{Department of Physics and Astronomy, Ohio University,
Athens, OH~45701, USA}

\date{\today}

\begin{abstract}

The characteristics of the hadron-to-quark first-order phase transition differ depending on whether charge neutrality is locally or globally fulfilled. In $\beta$-equilibrated matter, these two possibilities correspond to the Maxwell and Gibbs constructions. 
Recently, we presented a new framework in which a continuously-varying parameter allows one to describe a first-order phase transition in intermediate scenarios to the two extremes of fully local and fully global charge neutrality. In this work, we extend the previous framework to finite temperatures and out-of-$\beta$ equilibrium conditions, making it available for simulations of core-collapse supernovae and binary neutron star mergers. We investigate its impact on key thermodynamic quantities across a range of baryon densities, temperatures, and electron fractions. We find that when matter is not in $\beta$-equilibrium, the pressure in the mixed phase is not constant even for the case of fully-local charge neutrality. Moreover, we compute the thermal index using three different approaches, demonstrating that the finite-temperature extension of an equation of state using a constant thermal index can be ill-defined when applied to the mixed phase.
\end{abstract}

\maketitle

\section{Introduction}
\label{sec:intro}

Due to the running coupling constant of Quantum Chromodynamics (QCD), a comprehensive understanding of strong interactions cannot always be achieved through a simple perturbative analysis. Moreover, the relevant degrees of freedom of the theory depend on the ambient conditions. At high energy (i.e., high temperature or baryon density), weakly interacting quarks are the effective degrees of freedom. In contrast, quarks are confined within hadrons at low energy (i.e., low temperature and baryon density). Thus, a transition from hadronic to quark degrees of freedom is expected at some intermediate temperature and baryon density. 
While in the low baryon density and high-temperature regime, lattice QCD calculations found a crossover between the hadron gas and the quark-gluon plasma at $T\sim 155$ MeV~\cite{Borsanyi:2020fev}, no exact methods are currently available to study the phase transition at high density.
Such extreme density conditions are rare in the Universe and can only be reached in high-density astrophysical systems such as neutron stars (NSs). NSs have an average baryon density of $\nb\sim(2 - 3)\,n_0$, where $n_0\simeq 0.16$ fm$^{-3}$ is the nuclear saturation density. Since the typical temperature of isolated NSs ($\sim$ keV) is negligible with respect to their typical kinetic Fermi energy ($\sim$ 100 MeV), they can be considered $T = 0$ systems for all practical applications~\cite{Schaffner-Bielich:2020psc}. 
However, temperatures up to $\sim 100$ MeV can be reached in some high-energy astrophysical phenomena related to NSs, such as binary NS mergers (BNSMs)~\cite{Perego:2019adq} and protoneutron star (PNS) ~\cite{Prakash:1996xs} formation in the aftermath of core-collapse supernova explosions (CCSNe)~\cite{Fischer:2017zcr}.
 The cores of the most massive NSs, are likely composed of deconfined quark matter ~\cite{Annala:2019puf}. In principle, self-bound compact stars composed purely of deconfined quark matter may also exist (see, e.g.~\cite{Drago:2015cea,Drago:2015dea,Bombaci:2020vgw}).
For these reasons, NSs and some related high-energy astrophysical systems are interesting laboratories for studying dense matter in general and the hadron-to-quark transition in this regime in particular, both at zero and at finite temperatures.

The equation of state (EOS), namely, the relation between the system's thermodynamic quantities, is a key ingredient for computing the hydrostatic properties of isolated NSs, such as mass, radius, and tidal deformability, and for hydrodynamical simulations of CCSNe and BNSMs (see, e.g.~\cite{Oertel:2016bki,CompOSECoreTeam:2022ddl}). 
The main challenge of EOSs for astrophysical purposes is to describe a system in a vast range of baryon density $\nb\sim(10^{-8}-10) n_0$, temperature $T\sim(0-150)$ MeV, and net electron fraction $Y_e\sim0.01-0.6$~\cite{Oertel:2016bki}. Usually, the parameters of the models are fitted to laboratory data of the bulk nuclear properties at $\nb\sim n_0$ for cold and nearly symmetric matter. At the same time, behavior at higher densities is constrained by astrophysical data such as NS properties~\cite{Lattimer:2021emm}. For example, the observations of NSs with masses $\gtrsim 2\,\Msolar$~\cite{Antoniadis:2013pzd,Linares:2018ppq,Fonseca:2021wxt} have strongly constrained the stiffness of the EOS. In general, however, observations constrain the EOS only in small domains of thermodynamic variables, and a large amount of extrapolation is required to generalize models across the whole range of density, temperature, and isospin asymmetry. 
Moreover, current astrophysical observables cannot constrain matter composition in general and the presence of deconfined quarks in particular. Some crucial constraints on the composition of NSs may be present in the gravitational wave (GW) signals from the BNS post-merger phase (see, e.g.~\cite{Bauswein:2018bma,Prakash:2021wpz}). While current GW detectors only provide information on the inspiral part, third-generation detectors (Einstein Telescope, Cosmic Explorer) will eventually have sufficient sensitivity to study the post-merger dynamics~\cite{Prakash:2023afe}. This means that the new detectors will be able to provide information about the presence of deconfined quarks in astrophysical systems and the nature of the transition. Thus, the modeling of the hadron-to-quark transition in high-density astrophysical systems will be crucial in the following years for the correct interpretation of the GW data.

The nature of the hadron-to-quark transition under such conditions remains uncertain and could, in principle, manifest as either a first-order phase transition or a smooth crossover. Unlike a first-order phase transition which features sharp boundaries, a crossover transition occurs gradually, characterized by the mixing of two phases at the quantum level~\cite{Fukushima:2020cmk}. This behavior is typically modeled using schemes of smooth interpolation between hadronic and quark model EOSs~\cite{Masuda:2012kf,Baym:2017whm,Kapusta:2021ney}, or within the quarkyonic matter framework where a shell of nucleons resides atop a quark Fermi sea~\cite{McLerran:2018hbz}. 
In this work, we focus on the first-order phase transition scenario, which involves classical mixing and depends critically on the surface tension at the quark-hadron interface. Nevertheless, the magnitude of this surface tension is highly uncertain, ranging from a few to hundreds of MeV/fm$^2$ (see,  e.g.~\cite{Alford:2001zr,Mintz:2009ay,Palhares:2010be,Lugones:2018qgu,Fraga:2018cvr,Schmitt:2020tac,Ju:2021hoy}).

The description of first-order phase transitions for typical isolated NS conditions (i.e. cold, neutrino-less matter in $\beta$-equilibrium) is extensively discussed in the literature (see, e.g.~\cite{Han:2019bub}). 
The most commonly employed methods are the Maxwell (MC) and Gibbs constructions (GC)~\cite{Han:2019bub,Glendenning:1992vb}. 
In MC, only the baryon number is a globally conserved quantity, while charge neutrality is fulfilled locally (namely in the quark and hadronic phases separately). Meanwhile, both baryon number conservation and charge neutrality are achieved globally in GC. See~\cite{Hempel:2009vp} for a complete discussion regarding the role of conserved quantities in first-order phase transitions. Which of the two approaches better describes the mixed phase depends on the interplay between the surface tension and the Debye screening length~\cite{Heiselberg:1992dx,Voskresensky:2002hu}. 
A larger surface tension leads to bigger finite-size structures in the mixed phase. If their size is larger than the Debye length, then charge screening is efficient, and the bulk of the matter is nearly locally charge neutral. If, instead, the surface tension is small so that the typical structures are smaller than the Debye screening length, global charge neutrality is a more reasonable assumption~\cite{Voskresensky:2002hu}.
Thus, the two approaches are usually interpreted as the limiting cases in which the surface tension between the two phases is very large or negligible. In Appendix~\ref{apd:screen}, we show that the critical surface tension,
\begin{eqnarray}
    \sigma_{c} &=& 9.1 ~[\textrm{MeV/fm}^2] \left (\frac{n_e}{0.1  ~[\textrm{fm}^{-3}]}\right)
\end{eqnarray}
scales linearly with the global electron density, $n_e$.
A MC mixed phase is characterized by a constant pressure, while the pressure is a continuous function in the GC mixed phase.
These different features of MC and GC lead to different compact-object structures: in the MC case, the mixed phase has a vanishing extent since constant pressure cannot counteract the pull of gravity. 
In contrast, the GC supports a proper mixed-phase region in compact stars. The MC and GC neglect finite-size effects, and the system is considered in the bulk limit. However, a complete discussion should also consider finite-size effects~\cite{Heiselberg:1992dx} that play a relevant role in the formation of geometrical structures in the mixed phase (see, e.g.~\cite{Ju:2021hoy,Mariani:2023kdu}) and in the nucleation process (see, e.g.~\cite{Mintz:2009ay,Bombaci:2016xuj,Guerrini:2024gzu}). 
In~\cite{Constantinou:2023ged}, a new framework is presented in which charge neutrality is fulfilled partially locally and partially globally. This new approach describes first-order phase transitions in bulk in which, for NS matter, the MC and GC are just the two limiting cases. 

Generally, high-energy astrophysical simulations of CCSNe and BNSMs need EOSs with $\nb,Y_e,T$ as independent variables, namely finite-temperature EOSs in which equilibrium involving $\beta$-reactions is not assumed~\cite{CompOSECoreTeam:2022ddl}. 
Simulations using EOSs containing quark degrees of freedom are present in the literature. For example, in~\cite{Fischer:2017lag,Kuroda:2021eiv}, an EOS with a MC is used in a CCSN simulation. Here, the authors propose quark deconfinement as the mechanism that allows blue supergiant progenitors to explode after core collapse instead of collapsing into black holes. 
BNSM simulations using a MC description of the hadron-quark phase transition are presented in~\cite{Most:2019onn,Bauswein:2018bma}, while a GC approach has been employed in~\cite{Prakash:2021wpz} (see its introduction and references therein for a review of quark deconfinement in BNSMs). These works generally find that quark deconfinement qualitatively modifies the post-merger dynamics and the relative gravitational wave signal. However, reliable observables unambiguously related to deconfined quark degrees of freedom and to the nature of the phase transition have yet to be identified.

This work aims to generalize the framework presented in~\cite{Constantinou:2023ged} to the finite-temperature and out of $\beta-$equilibrium case to make it available for simulations of CCSNe and BNSMs. 
The advantage of using this approach in hydrodynamical simulations is the possibility of studying the impact of quark degrees of freedom, while controlling the local or global charge neutrality with only one continuous parameter in a thermodynamically consistent way.

We consider nucleons (i.e. protons and neutrons) as the only hadronic degrees of freedom, and discard contributions due to hyperons, deltas, or meson condensates. 
Moreover, neither superfluidity nor color-superconductivity~\cite{Alford:2007xm} are considered for both the nucleon and the quark sectors. 

We also neglect finite-size effects on the phase transition.
Finally, muons and neutrinos are not taken into account, since, despite their important role in high-energy astrophysical systems, their contributions are often included in simulations by post-processing techniques or explicit radiative transfer (see, e.g.~\cite{Radice:2021jtw}).

The organization of this work is as follows: In Sec.~\ref{sec:fwork}, the framework for the first-order phase transition is described. The EOS models for nucleons, quarks, electrons, and photons are given in Sec.~\ref{sec:eos}.
The numerical setup used to extend the framework to the finite temperature case is reported in Sec.~\ref{sec:numer}. A discussion of the results is presented in Sec.~\ref{sec:results}. A summary and the conclusions are presented in Sec.~\ref{sec:sum}

We will use natural units in which $\hbar=c=1$. 
We also use $k_B=1$ for temperature, specific entropy and specific heat.
%

\section{Framework}
\label{sec:fwork}

In a previous work~\cite{Constantinou:2023ged}, we discussed a thermodynamically consistent framework by which constructions for first-order hadron-to-quark transitions (other than the usual Maxwell and Gibbs) can be implemented at $T=0$. We argued that a quasi-permeable boundary between the two phases leads to phase mixing in its vicinity, whereas phase separation is maintained further away. In this picture, leptons in the former region are in contact with both phases, while the rest interact only with one or the other phase. As a result, charge neutrality is achieved partially locally and partially globally. 

To quantify the amount of mixing, we introduced the parameter $\eta \in [0,1]$ which measures the ratio of leptons involved in local charge neutrality (LCN) to the total number of leptons. Thus $\eta = 0$ corresponds to the Gibbs case of global charge neutrality (GCN) and $\eta = 1$ to the Maxwell case of LCN. The equilibrium state of the system is obtained by minimizing its total energy density $\ep$ with respect to the various particle fractions $y_i,~\eta$, and $\chi$ - the volume fraction of hadrons [with quarks having $(1-\chi)$]. Minimization is carried out under the constraints of baryon and lepton number conservation, in addition to charge neutrality.

A complementary view is to assume the formation of pure-phase lumps in the coexistence region. Depending on the size of these lumps (relative to the Debye screening length), leptons will behave as if embedded in a phase-separated, partially mixed, or completely mixed environment. In such a perspective, the parameter $\eta$ may be thought of as a crude measure for the size of extended structures in the mixed phase.

In the present, we deal with the extension of this framework to finite temperature. This is a fairly straightforward task requiring the minimization of the total \textit{free} energy density $\fed$ of the system (as opposed to $\ep$), which must also include contributions from antiparticles. Explicitly, 
\beq
\fed &=& \chi\left[\fed_n + \fed_p+ \eta(\fed_{\eN} + \fed_{\bareN})\right] 
\nonumber \\
&+& (1-\chi)\left[\fed_u + \fed_{\bar u} + \fed_d + \fed_{\bar d} \right. 
\nonumber \\
&+& \left. \fed_s + \fed_{\bar s} + \eta(\fed_{\eQ} + \fed_{\bareQ})\right] \nonumber \\
&+& (1-\eta)\left(\fed_{\eG} + \fed_{\bareG}\right)
\label{eqn:fden}
\eeq
where the barred quantities represent antiparticles, and the subscripts $\eN,~\eQ,~\eG$ refer to the electrons associated with nucleons, quarks, and a mixture thereof, respectively. 
For $T\lesssim 100$ MeV relevant for CCSNe and BNSMs, the antinucleons can be safely ignored. For simplicity, in this work we also neglect muons. 

At finite temperature, the constraints of baryon and lepton number conservation are modified to 
\beq
1 &=& \chi \,(y_n + y_p) + (1-\chi)(Y_u + Y_d + Y_s)/3 \label{eqn:baryoncons}  \\
0 &=& Y_e - \chi\eta Y_{\eN} - (1-\chi)\eta Y_{\eQ} - (1-\eta)Y_{\eG}\label{eqn:leptoncons}
\eeq
where $Y_i \equiv y_i - y_{\bar i}$ are the \textit{net} particle fractions. Correspondingly, the charge neutrality conditions are given by
\beq
0 &=& y_p - Y_{\eN}  \label{eqn:chargeneutH}\\ 
0 &=& (2Y_u - Y_d - Y_s)/3 - Y_{\eQ}  \label{eqn:chargeneutQ}\\ 
0 &=& \chi y_p + (1-\chi)(2Y_u - Y_d - Y_s)/3 - Y_{\eG} ~.\label{eqn:chargeneutglob}
\eeq
Following the procedure delineated in Ref.~\cite{Constantinou:2023ged}, one obtains a set of equilibrium conditions which is nearly identical to the $T=0$ case:
\beq
\mu_n &=& \mu_u + 2\mu_d 
\label{eqn:strong1}\\
\mu_p &=& 2\mu_u + \mu_d - \eta(\mu_{\eN} - \mu_{\eQ})  
\label{eqn:strong2}\\
\mu_d &=& \mu_u + \eta \mu_{\eQ} + (1-\eta)\mu_{\eG}  \label{eqn:beta}\\
\mu_d &=& \mu_s    \label{eqn:qweak}\\
\fed_{\eG,\bareG} &=& \chi \fed_{\eN,\bareN} + (1-\chi)\fed_{\eQ,\bareQ}  \label{eqn:mineta}\\
P_{N} &+& \eta P_{\eN,\bareN} = P_{Q,\bar{Q}} + \eta P_{\eQ,\bareQ} \label{eqn:meceq}
\eeq

Note, however, that the free energy densities and pressures appearing in the last two equations now contain antiparticle terms, in addition to particle ones. Furthermore, finite-$T$ equilibrium requires the usual relation between the chemical potentials of particles and antiparticles of a given species:
\be
\mu_i = -\mu_{\bar i} ~~;~~ i=u,d,s,\mbox{$\eN,\eQ,\eG$} ~.
\ee

We point out that, in general, $\beta$ reactions do not equilibrate in astrophysical events such as CCSNe and BNSMs and, therefore, the corresponding condition, \Eqn{eqn:beta}, does not apply; the effect being that the net lepton fraction $Y_e$ remains an independent variable. On the other hand, dynamical timescales pertaining to the aforementioned phenomena are much longer than those of flavor-changing weak interactions (ms vs.~$\mu$s)~\cite{Brown:1992ib,Oertel:2016bki}, which means that \Eqn{eqn:qweak} must be enforced. 
This is not the case for systems where strangeness is conserved as, for example, in heavy-ion collisions. 
Note that \Eqn{eqn:mineta} is derived from minimization with respect to $\eta$; although we include it for completeness, we will not use it for the sake of keeping $\eta$ a free variable to explore the effects of the changing surface tension.

\section{Equation of state}
\label{sec:eos}
We are interested in the finite-temperature properties of the ZL~\cite{Zhao:2020dvu,Constantinou:2021hba} and vMIT~\cite{Klahn:2015mfa, Gomes:2018eiv} models for which the single-particle energy spectrum (quasi-particle energy) of species $i$ has the functional form 
\beq
\epsilon_{k_i} &=& (k_i^2 + m_i^2)^{1/2} + U_i(\nb,\{y_i\}) \nonumber \\
&\equiv& E_{k_i} + U_i(\nb,\{y_i\})~.
\eeq
Microscopically, such models can arise from vector meson exchange at the mean-field level (relativistic Hartree approximation). Scalar mesons introduce additional complications because they couple to source masses which, therefore, become dependent upon the properties of the medium (Dirac masses). We caution, however, that ZL is a schematic model without a microscopic basis. Its potential contains non-integer powers of the density which cannot be obtained from vector meson exchange.

Given an energy density functional $\ep(\nb,\{\tau_i\},\{y_i\})$, where $\tau_i$ is the kinetic energy density of species $i$ 
\be
\tau_i = \gamma_i\int_0^{\infty} \frac{d^3k_i}{(2\pi)^3}~\frac{E_{k_i}}
{e^{(\epsilon_{k_i}-\mu_i)/T}+1}~, 
\ee
and $y_i$ its concentration
\be
y_i  =  \frac{\gamma_i}{\nb}\int_0^{\infty}\frac{d^3k_i}{(2\pi)^3}
\frac{1}{e^{(\epsilon_{k_i}-\mu_i)/T}+1}. 
\label{eqn:yi}
\ee
The spectrum is obtained by functional differentiation according to 
\be
\epsilon_{k_i} = E_{k_i}~\frac{\partial \ep}{\partial \tau_i} 
+ \frac{1}{\nb}\frac{\partial \ep}{\partial y_i} ~.
\ee
Above, $\gamma_i$ is a degeneracy factor (spin, isospin, color, etc).

As a consequence of $U_i$ being momentum-independent for the models under consideration, the exponents in the denominators of $\tau_i$ and $y_i$ can be written as 
\be
\frac{\epsilon_{k_i}-\mu_i}{T} = \frac{E_{k_i}-\nu_i}{T}
\ee
where $\nu_i = \mu_i - U_i$ is the solution of \Eqn{eqn:yi}. 
Thus the integrals are reduced to those of a relativistic ideal gas. 

Correspondingly, the energy density of species $i$ becomes
\beq
\ep_i &=& \gamma_i\int_0^{\infty} \frac{d^3k_i}{(2\pi)^3}~\frac{\epsilon_{k_i}}
{e^{(\epsilon_{k_i}-\mu_i)/T}+1}  \nonumber  \\
&=& \gamma_i\int_0^{\infty} \frac{d^3k_i}{(2\pi)^3}~\frac{E_{k_i}}
{e^{(E_{k_i}-\nu_i)/T}+1} + V_{\ep_i}  \nonumber \\
&\equiv& I_{\ep_i} + V_{\ep_i}
\eeq
where $I_{\ep_i}$ is the integral term and $V_{\ep_i}$ is the potential contribution to $\ep_i$, identical to the $T=0$ case.

The pressure is obtained from 
\beq
P_i &=& \nb \left.\frac{\partial \ep_i}{\partial \nb}\right|_{\{y_i\},T} 
\nonumber \\
&=& \gamma_i\int_0^{\infty} \frac{d^3k_i}{(2\pi)^3}~k_i
\frac{\partial E_{k_i}}{\partial k_i}\frac{1}{e^{(E_{k_i}-\nu_i)/T}+1} 
+ \left.\nb\frac{\partial V_{\ep_i}}{\partial \nb}\right|_{\{y_i\}}  
\nonumber \\
&=& \gamma_i\int_0^{\infty} \frac{d^3k_i}{(2\pi)^3}~
\frac{k_i^2}{E_{k_i}}\frac{1}{e^{(E_{k_i}-\nu_i)/T}+1} 
+ \left.\nb\frac{\partial V_{\ep_i}}{\partial \nb}\right|_{\{y_i\}}
\nonumber \\
&\equiv & I_{P_i} + V_{P_i}
\eeq
where, similar to before, $I_{P_i}$ refers to the integral and $V_{P_i}$ is the $T=0$ potential component of the pressure.

In the ZL model, $V_{\ep_H} = V_{\ep_n} + V_{\ep_p}$ is given by 
\beq
V_{\ep_H} &=& 4 \nb^2 y_n y_p \left\{\frac{a_0}{\nsat} 
+\frac{b_0}{\nsat^{\gamma}} [\nb(y_n + y_p)]^{\gamma - 1}\right\}  
\nonumber \\
&+& \nb^2 (y_n - y_p)^2\left\{\frac{a_1}{\nsat} 
+ \frac{b_1}{\nsat^{\gamma_1}}[\nb(y_n + y_p)]^{\gamma_1-1}\right\}
\nonumber \\
\eeq
with our choices for the parameters appearing in this expression, along with those for vMIT, shown in Table~\ref{tab:pars}. Note that for the ZL model, we use the ZLA parametrization as in \cite{Constantinou:2023ged}. 
\begin{table}[h]
\caption{
Parameter sets used in the present work. Different choices of $B$ can be made to tune the baryon density range of the mixed phase. Units of $c=1$ are employed. } 
\begin{center} 
\begin{tabular}{ccrc}
\hline
\hline
Model       & Parameter  & Value       & Units \\ \hline
            &  $n_0$     & 0.16    & fm$^{-3}$     \\
            &  $a_0$     & -96.64    & MeV     \\
            &  $b_0$     &  58.85    & MeV     \\
  ZLA        &  $\gamma$  &  1.40     &         \\
            &  $a_1$     & -26.06    & MeV     \\
            &  $b_1$     &  7.34     & MeV     \\
            & $\gamma_1$ &  2.45     &         \\
\hline 
            & $m_u$      &  5.0     & MeV     \\
            & $m_d$      &  7.0     & MeV     \\
  vMIT      & $m_s$      &  150.0   & MeV     \\
            & $a$        & 0.20     & fm$^2$  \\
            & $B^{1/4}$  & 180   & MeV     \\
\hline
            & {$\hbar(c)$}    & 197.3     & {MeV~fm}    \\
 Constants  & $m_e$        & 0.511     & {MeV} \\
\hline \hline
\end{tabular}
\end{center}
\label{tab:pars}
\end{table}

The corresponding energy density, pressure, and chemical potentials are
\beq
\ep_H &=& \sum_{h=n,p}I_{\ep_h} + V_{\ep_H}  \\
P_H &=& \sum_{h=n,p}I_{P_h} 
+ \left.\nb\frac{\partial V_{\ep_H}}{\partial \nb}\right|_{\{y_h\}}  \\
\mu_{h_1} &=& \nu_{h_1} + \frac{1}{\nb}
\left.\frac{\partial V_{\ep_H}}{\partial y_{h_1}}\right|_{\nb,y_{h_2}}~.
\eeq 

In the quark sector, antiquark terms must now be incorporated in the potential as well as in the kinetic energy: 

\beq
V_{\ep_Q} &=& \sum_{\substack{q=u,d,s,\\~~{\bar u},{\bar d},{\bar s}}}V_{\ep_q}   \nonumber \\
&=& \frac{1}{2}a
\left[\nb \left(\sum_{\substack{q=u,d,s}}Y_q\right)\right]^2+B~,
\eeq
where the parameter $a$ controls the strength of the vector repulsion between quarks, while $B$ is the bag constant, representing the energy density difference between the non-perturbative and perturbative vacuum, acting effectively as a confining pressure (see \cite{Klahn:2015mfa, Gomes:2018eiv} for the details).
Therefore, 
\beq
\ep_Q &=& \sum_{{\substack{q=u,d,s,\\~~{\bar u},{\bar d},{\bar s}}}}I_{\ep_q} + V_{\ep_Q}  \\
P_Q &=& \sum_{{\substack{q=u,d,s,\\~~{\bar u},{\bar d},{\bar s}}}}I_{P_q} 
+ \left.\nb\frac{\partial V_{\ep_Q}}{\partial \nb}\right|_{\{y_q\}}  \\
\mu_{q_i} &=& \nu_{q_i} + \frac{1}{\nb}
\left.\frac{\partial V_{\ep_Q}}{\partial y_{q_i}}\right|_{\nb,y_{q_{j\ne i}}}~.
\eeq 

Electrons and positrons are, per usual, treated as noninteracting, relativistic fermions:
\beq
\ep_L &=& \sum_{l=e,\bar e}I_{\ep_l}  \\
P_L &=& \sum_{l=e,\bar e}I_{P_l}  \\
\mu_{l} &=& \nu_{l} ~.
\eeq 
Photons are treated as an ideal boson gas with zero chemical potential (i.e. as blackbody radiation): 
\beq
\ep_{\gamma}&=&\frac{\pi^2}{15}T^4\\
P_{\gamma}&=&\frac{\ep_{\gamma}}{3}\\
s_{\gamma}&=&\frac{4\pi^2}{45}T^3\\
c_{V\gamma}&=&\frac{4\pi^2}{15}T^3\\
c_{P\gamma}&=&\frac{4}{3}c_{V\gamma}
\eeq
Other thermodynamic quantities of interest such as the entropy density and the free energy density follow from standard relations:
\beq
s &=& \frac{1}{T}\left(\ep + P + \nb\sum_i \mu_iy_i \right)\\
\fed &=& \ep - Ts ~.
\eeq
%

\section{Numerics}
\label{sec:numer}
In the interests of thermodynamic consistency and computational efficiency, we employ the JEL method~\cite{Johns:1996ht} for the numerical evaluation of the thermodynamic integrals. In this approach, the concentration $y_i$, the kinetic energy density $I_{\ep_i}$, and the kinetic pressure $I_{P_i}$ of particle species $i$ are expressed algebraically in terms of the  mass, the temperature, and a parameter $f_i$ related to the chemical potential: 
\beq
y_i &=& \frac{\gamma}{\pi^2}\frac{m_i^3}{\nb}
  \frac{f_ig_i^{3/2}(1+g_i)^{3/2}}{(1+f_i)^{M+1/2}(1+g_i)^N(1+f_{i}/a)^{1/2}} \nonumber  \\
  &\times& \sum_{m=0}^M\sum_{n=0}^Np_{mn}f_i^mg_i^n
  \left[1+m+\left(\frac{1}{4}+\frac{n}{2}-M\right)\frac{f_i}{1+f_i}\right. 
  \nonumber  \\ 
  &+& \left.\left(\frac{3}{4}-\frac{N}{2}\right)\frac{f_ig_i}{(1+f_i)(1+g_i)}\right] \label{eq:yiJEL}\\
I_{\ep_i} &-& m_i \nb y_i \equiv I_{\ep_i}^-   \\
&=& \frac{\gamma m_i^4}{\pi^2} \frac{f_ig_i^{5/2}(1+g_i)^{3/2}}{(1+f_i)^{M+1}(1+g_i)^N}
              \sum_{m=0}^M\sum_{n=0}^Np_{mn}f_i^mg_i^n    \nonumber \\
       &\times& \left[\frac{3}{2}+n+\left(\frac{3}{2}-N\right)\frac{g_i}{1+g_i}\right]  \label{eq:IeiJEL}\\
I_{P_i} &=& \frac{\gamma m_i^4}{\pi^2} \frac{f_ig_i^{5/2}(1+g_i)^{3/2}}{(1+f_i)^{M+1}(1+g_i)^N}
              \sum_{m=0}^M\sum_{n=0}^Np_{mn}f_i^mg_i^n ~.\label{eq:IPiJEL}
\eeq
where $I_{\ep_i}^-$ is the kinetic energy density exclusive of the rest-mass density and $g_i \equiv \frac{T}{m_i}(1+f_i)^{1/2}\equiv t_i(1+f_i)^{1/2}$.
In these expressions $\gamma = 1$ for nucleons and leptons and 3 for quarks, while $f_i$ is connected to the fugacity $\psi_i$ and the (free) chemical potential $\nu_i$ according to
\begin{equation}
\psi_i = \frac{\nu_i-m_i}{T}=2(1+f_i/a)^{1/2}+\ln\left[\frac{(1+f_i/a)^{1/2}-1}{(1+f_i/a)^{1/2}+1}\right] ~.
\end{equation}
We emphasize that $m_i$ and $\nb$ exterior to the sums in Eqs. (\ref{eq:yiJEL},\ref{eq:IeiJEL},\ref{eq:IPiJEL}) correspond to particle mass and baryon density, whereas $m$ and $n$ inside the sums denote binomial coefficients.

The coefficients $p_{mn}$ for $M=N=3$ and $a=0.433$ are displayed in Table~\ref{tab:jel}. \\ 
\begin{table}[h]
\begin{center}
\begin{tabular}{lllll}
\hline \hline
$p_{mn}$ & $n=0$ & $n=1$ & $n=2$ & $n=3$ \\
\hline
$m=0$ & 5.34689 & 18.0517 & 21.3422 & 8.53240 \\
$m=1$ & 16.8441 & 55.7051 & 63.6901 & 24.6213 \\
$m=2$ & 17.4708 & 56.3902 & 62.1319 & 23.2602 \\
$m=3$ & 6.07364 & 18.9992 & 20.0285 & 7.11153 \\
\hline  \hline
\end{tabular}
\caption[JEL Coefficients]{JEL coefficients $p_{mn}$ for $M=N=3$ and $a=0.433$.}
\label{tab:jel}
\end{center}
\end{table}  
%

\subsection{Specific heat at constant volume}
The specific heat at constant volume of a multicomponent system is given by 
\be
C_V = \left.\frac{\p E}{\p T}\right|_{\nb,\{y_i\}}
=\left.\frac{\p \sum_i y_i E_i}{\p T}\right|_{\nb,\{y_i\}} 
=\sum_i y_iC_{Vi}~\label{eq:Cv1},
\ee
where $E$ is the total energy \textit{per baryon}, $E_i$ the energy per particle of the particle species $i$, and $C_{V_i}$ the specific heat of particle species $i$. 

In the mixed phase, as described by \Eqn{eqn:fden}, the \Eqn{eq:Cv1} must be modified to 
\be
C_V = \sum_i r_i y_i C_{Vi}
\ee
with $r_i = \chi$ for $i=(n,p)$, $\chi \eta$ for $i=(\eN,\bareN)$, 
$(1-\chi)$ for $i=(q,\bar{q})$, $(1-\chi) \eta$ for $i=(\eQ,\bareQ)$, and 
$(1-\eta)$ for $i=(\eG,\bareG)$. It is understood that, here, $y_i$ refers to pure-phase fractions and $y_i^* \equiv r_i y_i$ to concentrations in the mixed phase.

For a pure system with only vector interactions at the mean-field level (such as the ZL and vMIT models under consideration) 
\beq
C_{V_i} &=& \left.\frac{\p}{\p T}\left(\frac{I_{\ep_i}+V_{\ep_i}}{y_i \nb }\right)
        \right|_{\nb,\{y_i\}} \nonumber \\
\Rightarrow y_i~C_{V_i} &=&\frac{1}{\nb}
\left.\frac{\p \sum_i I_{\ep_i}^-}{\p T}\right|_{\nb,\{y_i\}}~,
\eeq
being that $V_{\ep}$ is independent of the temperature. 
On the other hand, in the presence of scalar interactions or higher-order many-body effects,  $m\rightarrow m^*(n,T)$ and thus 
$\p m^*/\p T \ne 0$ (among other complications). 

Given these facts, the following calculation proceeds in a single-component context; that is, the subscript $i$ denoting particle species is suppressed. In order to connect with the JEL approach, we will initially think of $I_{\ep}^-$ as a function of the density $\nb$, the concentration $y$, the temperature $T$, and the free chemical potential $\nu$: 
\beq
I_{\ep}^- &=& I_{\ep}^-[\nb,y,T,\nu(\nb,y,T)]   \\
\Rightarrow y~C_V &=& \frac{1}{\nb}\left(\left.\frac{\p I_{\ep}^-}{\p T}\right|_{\nb,y,\nu} + 
\left.\frac{\p I_{\ep}^-}{\p \nu}\right|_{\nb,y,T}\left.\frac{\p \nu}{\p T}\right|_{\nb,y}\right)~. \nonumber \\
\eeq
Subsequently, we switch to the JEL variables $\psi(\nu,T)$ and $t(\nu,T)$ such that 
\beq
y~C_V &=& \frac{1}{\nb}\left[\left.\frac{\p I_{\ep}^-}{\p \psi}\right|_t
        \left.\frac{\p \psi}{\p T}\right|_\nu
    +   \left.\frac{\p I_{\ep}^-}{\p t}\right|_\psi
        \left.\frac{\p t}{\p T}\right|_\nu  \right. \nonumber \\
  &+&   \left. \left.\frac{\p \nu}{\p T}\right|_{\nb,y}
        \left(\left.\frac{\p I_{\ep}^-}{\p \psi}\right|_t
        \left.\frac{\p \psi}{\p \nu}\right|_T
   +    \left.\frac{\p I_{\ep}^-}{\p t}\right|_\psi
        \left.\frac{\p t}{\p \nu}\right|_T\right)\right] \nonumber \\
\eeq
A fixed $y = y[\nb,T,\nu(\nb,y,T)]$ means that 
\beq
\frac{dy}{dT} &=& 0 = \left.\frac{\p y}{\p T}\right|_{\nb,\nu} 
                  + \left.\frac{\p y}{\p \nu}\right|_{\nb,T}
                    \left.\frac{\p \nu}{\p T}\right|_{\nb,y}  \\
\Rightarrow \left.\frac{\p \nu}{\p T}\right|_{\nb,y} &=& 
 -\frac{\p y/\p T|_{\nb,\nu}}{\p y/\p \nu|_{\nb,T}}~.
\eeq
For the derivatives of $y$, we begin by writing $y= y[\nb,\psi(\nu,T),t(\nu,T)]$ and therefore 
\beq
\left.\frac{\p y}{\p T}\right|_{\nb,\nu} &=&
    \left.\frac{\p y}{\p \psi}\right|_{\nb,t} 
    \left.\frac{\p \psi}{\p T}\right|_\nu  +
    \left.\frac{\p y}{\p t}\right|_{\nb,\psi} 
    \left.\frac{\p t}{\p T}\right|_\nu   
    \label{dydtnu}  \\
\left.\frac{\p y}{\p \nu}\right|_{\nb,T} &=&
    \left.\frac{\p y}{\p \psi}\right|_{\nb,t} 
    \left.\frac{\p \psi}{\p \nu}\right|_T  +
    \left.\frac{\p y}{\p t}\right|_{\nb,\psi} 
    \left.\frac{\p t}{\p \nu}\right|_T 
    \label{dydnut} ~.
\eeq
The definitions $\psi = \frac{\nu-m}{T}$ and $t=T/m$ imply that 
\beq
\left.\frac{\p \psi}{\p \nu}\right|_T = \frac{1}{T} ~~&;&~~
\left.\frac{\p \psi}{\p T}\right|_\nu = -\frac{\psi}{T}   \\
\left.\frac{\p t}{\p \nu}\right|_T = 0 ~~&;&~~ 
\left.\frac{\p t}{\p T}\right|_\nu = \frac{1}{m}
\eeq
and thus Eqs.~(\ref{dydtnu})-(\ref{dydnut}) become
\beq
\left.\frac{\p y}{\p T}\right|_{\nb,\nu} 
 &=& -\frac{\psi}{T}\left.\frac{\p y}{\p \psi}\right|_{\nb,t}
   + \frac{1}{m} \left.\frac{\p y}{\p t}\right|_{\nb,\psi}  \\
\left.\frac{\p y}{\p \nu}\right|_{\nb,T}  
 &=& \frac{1}{T} \left.\frac{\p y}{\p \psi}\right|_{\nb,t}  \\
\Rightarrow \left.\frac{\p \nu}{\p T}\right|_{\nb,y} 
 &=& \psi - t~\frac{\p y/\p t|_{\nb,\psi}}{\p y/\p \psi|_{\nb,t}}~.
\eeq
Putting everything together and restoring the particle subscript, the specific heat at constant volume of species $i$ is given by
\beq
y_i~C_{V_i} &=& \frac{1}{\nb}\left[\left.
\frac{\p I_{\ep_i}^-}{\p \psi_i}\right|_{t_i} \left(\frac{-\psi_i}{T}\right)
+ \left.\frac{\p I_{\ep_i}^-}{\p t_i}\right|_{\psi_i} 
  \left(\frac{1}{m_i}\right)\right. \nonumber \\
&+& \left. \left(\psi_i 
- t_i~\frac{\p y_i/\p t_i|_{\nb,\psi_i}}{\p y/\p \psi_i|_{\nb,t_i}}\right)
\left(\left.
\frac{\p I_{\ep_i}^-}{\p \psi_i}\right|_{t_i} \frac{1}{T}\right)\right]   \nonumber  \\
&=& \frac{1}{m_i\nb}\left(\left.\frac{\p I_{\ep_i}^-}{\p t_i}\right|_{\psi_i} 
- \frac{\p y_i/\p t_i|_{\nb,\psi_i}}{\p y_i/\p \psi_i|_{\nb,t_i}}
  \left.\frac{\p I_{\ep_i}^-}{\p \psi_i}\right|_{t_i}\right) ~. 
\label{eqn:yicvi}  
\nonumber \\
\eeq        
The specific heat of a multicomponent pure phase will be a sum of several terms such as \Eqn{eqn:yicvi}, 
each corresponding to a particle species present in the system. In a mixed phase, all these terms must be weighted with the appropriate factors $r_i$.

\subsection{Specific heat at constant pressure}
The specific heat at constant pressure is given by
\be
C_P = \left.\frac{\p H}{\p T}\right|_{P,\{y_i\}} 
\ee
where $H$ is the enthalpy per particle. 
A Legendre transform from the variables $(P,\{y_i\},T)$ to $(\nb,\{y_i\},T)$ results in~\cite{Landau:1980mil}
\be
C_P = C_V+\frac{T}{\nb^2}\frac{\left(\p P/\p T|_{\nb,\{y_i\}}\right)^2}
{\p P/\p \nb|_{T,\{y_i\}}} ~. 
\label{eqn:cpnt}
\ee
The \textit{total} pressure is $P = \sum_i r_i P_i$ and therefore
\be
\left.\frac{\p P}{\p T}\right|_{\nb,\{y_i\}} = 
\sum_i r_i \left.\frac{\p I_{P_i}}{\p T}\right|_{\nb,\{y_i\}} ~.
\ee
The calculation of $\p I_{P_i}/\p T$ mirrors that of $\p I_{\ep_i}^-/\p T$ from the previous section with the outcome 
\be
\left.\frac{\p I_{P_i}}{\p T}\right|_{\nb,y_i} =
\frac{1}{m_i}\left(\left.\frac{\p I_{P_i}}{\p t_i}\right|_{\psi_i} 
      - \frac{\p y_i/\p t_i|_{\nb,\psi_i}}{\p y_i/\p \psi_i|_{\nb,t_i}}
      \left.\frac{\p I_{P_i}}{\p \psi_i}\right|_{t_i}\right)  ~.  
\ee  
For the derivative in the denominator of \Eqn{eqn:cpnt} 
we have
\be
\left.\frac{\p P}{\p \nb}\right|_{T,\{y_i\}}= 
\sum_i r_i \left(\left.\frac{\p I_{P_i}}{\p \nb}\right|_{T,y_i}
+ \left.\frac{\p V_{P_i}}{\p \nb}\right|_{T,y_i}\right) ~.
\ee
The potential derivative is specific to the models used and can be performed directly. Thus we focus our attention on $\p I_{P_i}/\p \nb$, once again suppressing the particle subscript until the final step.
 
We begin by writing $I_P = I_P\,[\nb,y,T,\nu(\nb,y,T)]$, which implies that 
\be
\left.\frac{\p I_P}{\p \nb}\right|_{T,y} = 
\left.\frac{\p I_P}{\p T}\right|_{\nb,y,\nu} + 
\left.\frac{\p I_P}{\p \nu}\right|_{\nb,y,T}\left.\frac{\p \nu}{\p T}\right|_{\nb,y} ~.
\label{eqn:dipdnb}
\ee
Shifting to the JEL variables $\psi$ and $t$ such that 
$I_P = I_P[\psi(\nu,T),t(\nu,t)]$, 
\Eqn{eqn:dipdnb} 
becomes
\beq
\left.\frac{\p I_P}{\p \nb}\right|_{T,y} &=& 
        \left.\frac{\p I_P}{\p \psi}\right|_t
        \left.\frac{\p \psi}{\p \nb}\right|_{\nu,T}
    +   \left.\frac{\p I_P}{\p t}\right|_\psi
        \left.\frac{\p t}{\p \nb}\right|_{\nu,T}   \nonumber \\
  &+&   \left.\frac{\p \nu}{\p \nb}\right|_{y,T}
        \left(\left.\frac{\p I_P}{\p \psi}\right|_t
        \left.\frac{\p \psi}{\p \nu}\right|_T
   +    \left.\frac{\p I_P}{\p t}\right|_\psi
        \left.\frac{\p t}{\p \nu}\right|_T\right) ~. \nonumber \\
\eeq
Given the definitions of $\psi$ and $t$, their only nonzero derivative with respect to $\nb$ or $\nu$ is $\p \psi/\p \nu|_T = 1/T$ and thus
\be
\left.\frac{\p I_P}{\p \nb}\right|_{T,y} = 
        \left.\frac{\p \nu}{\p \nb}\right|_{y,T}
        \left.\frac{\p I_P}{\p \psi}\right|_t
        \frac{1}{T} ~. \nonumber \\
\ee

Next, we must determine $\p \nu/\p \nb|_{y,T}$. For this, we express the concentration as $y = y[\nb,T,\nu(\nb,y,T)]$, set its total derivative with respect to $\nb$ equal to zero, and solve for $\p \nu/\p \nb|_{y,T}$:
\beq
\frac{dy}{d\nb} &=& 0 = \left.\frac{\p y}{\p \nb}\right|_{T,\nu} 
                  + \left.\frac{\p y}{\p \nu}\right|_{\nb,T}
                    \left.\frac{\p \nu}{\p \nb}\right|_{y,T}  \\
\Rightarrow \left.\frac{\p \nu}{\p \nb}\right|_{y,T} &=& 
 -\frac{\p y/\p \nb|_{T,\nu}}{\p y/\p \nu|_{\nb,T}}~.
\eeq

Finally, we let $y = y[\nb,\psi(\nu,T),t(\nu,t)]$ from which it follows that
\beq
\left.\frac{\p y}{\p \nb}\right|_{T,\nu} &=&
\left.\frac{\p y}{\p \nb}\right|_{\psi,t}   \\
\left.\frac{\p y}{\p \nu}\right|_{\nb,T} &=& 
\left.\frac{\p y}{\p \psi}\right|_{\nb,t}
\left.\frac{\p \psi}{\p \nu}\right|_T
+\left.\frac{\p y}{\p t}\right|_{\nb,\psi}
\left.\frac{\p t}{\p \nu}\right|_T
\nonumber \\
&=& \left.\frac{\p y}{\p \psi}\right|_{\nb,t}\frac{1}{T} \\
\Rightarrow \left.\frac{\p \nu}{\p \nb}\right|_{y,T} &=&
-T\frac{\p y/\p \nb|_{\psi,t}}{\p y/\p \psi|_{\nb,t}}~.
\eeq
Thus 
\be
\left.\frac{\p I_{P_i}}{\p \nb}\right|_{T,y_i} = 
       - \left.\frac{\p I_{P_i}}{\p \psi_i}\right|_{t_i}
       \frac{\p y_i/\p \nb|_{\psi_i,t_i}}{\p y_i/\p \psi_i|_{\nb,t_i}}
\ee
where the particle index $i$ has been recovered. 
\subsection{Adiabatic sound speed}
\label{sec:adss}
The adiabatic sound speed squared is defined as 
\beq
\left(\frac{\cad}{c}\right)^2 &=& 
       \left.\frac{\p P}{\p \ep}\right|_{S,\{y_i\}} 
       =\left.\frac{\p P}{\p \nb}\right|_{S,\{y_i\}} 
        \left(\left.\frac{\p \ep}{\p \nb}\right|_{S,\{y_i\}}\right)^{-1} 
        \label{cad1} \\
&=&  \frac{\nb}{\ep+P-Ts}\left.\frac{\p P}{\p \nb}\right|_{S,\{y_i\}} ~.        
\eeq
A Legendre transform~\cite{Landau:1980mil} of $\partial P/\partial \nb|_{S,\{y_i\}}$ to the variables $(\nb,\{y_i\},T)$ leads to 
\be
\left.\frac{\p P}{\p \nb}\right|_{S,\{y_i\}} =
\frac{C_P}{C_V}\left.\frac{\p P}{\p \nb}\right|_{T,\{y_i\}} ~.
\ee
Thus, the adiabatic sound speed squared can be expressed as 
\be
\left(\frac{\cad}{c}\right)^2 = 
\frac{C_P}{C_V}\frac{\nb}{(\ep+P-Ts)}\left.\frac{\p P}{\p \nb}\right|_{T,\{y_i\}}  
\label{cad}
\ee                                      
with all necessary ingredients for its numerical evaluation contained in previous sections and Appendix~\ref{apd:JEL}.

\subsubsection*{Slow vs.\ fast perturbations}
In writing Eqs.~\eqn{cad1}-\eqn{cad} in this form, it is implicitly assumed that the propagation of fluctuations due to some perturbation experienced by the system takes place over a timescale that is too short for the various reactions to reach equilibrium. Therefore, the corresponding chemical potential relations do not apply and all particle fractions remain free variables which, consequently, must be held constant when performing the partial derivatives; equilibration conditions are imposed only afterwards. 
\\
\indent Conversely, if this propagation is slow enough, all chemical equilibrium conditions, except the one for $\beta$ reactions, are enforced prior to taking the derivatives with attendant modifications to the adiabatic sound speed and other thermodynamic quantities of interest that depend on second derivatives of the free energy such as the specific heats at constant volume and pressure. We note that pressure equilibrium is also enforced a posteriori because the hadron volume fraction 
$\chi$ (the minimization with respect to which generates pressure equilibrium) must remain free in order to ensure the continuity of second- and higher-order derivatives in the mixed phase being that, in chemical equilibrium, $\p \chi/\p \nb \ne 0$.
\\
\indent First derivatives are not affected by these considerations. To see this, let $f_1=f(\nb,Y_e,Y)$, where, in this context,  $(\nb,Y_e,Y)$ are generic variables of the arbitrary function $f$.
Then
\beq
df_1 = \left.\frac{\p f}{\p \nb}\xy d\nb + \left.\frac{\p f}{\p Y_e}\ny dY_e + \left.\frac{\p f}{\p Y}\nx dY   \nn
\eeq
which implies
\beq
\left.\frac{\p f_1}{\p \nb}\xy &=& \left.\frac{\p f}{\p \nb}\xy  \nn \\
\left.\frac{\p f_1}{\p Y_e}\ny &=& \left.\frac{\p f}{\p Y_e}\ny  \nn \\
\left.\frac{\p f_1}{\p Y}\nx &=& \left.\frac{\p f}{\p Y}\nx ~. \nn 
\eeq
Let $\ye = \ye(\nb,Y_e)$ be such that $\left.\frac{\p f}{\p \ye}\nx = 0$ \\ and $f_2=f[\nb,Y_e,\ye(\nb,Y_e)]$. The differential of $f_2$ is 
\beq
df_2 &=& \left.\frac{\p f}{\p \nb}\xye d\nb + \left.\frac{\p f}{\p Y_e}\nye \nn dY_e\\
&+& \left.\frac{\p f}{\p \ye}\nx 
\left(\left.\frac{\p \ye}{\p \nb}\x d\nb + \left.\frac{\p \ye}{\p Y_e}\n dY_e \right)  \nn  \\
&=& d\nb\left(\left.\frac{\p f}{\p \nb}\xye + \left.\frac{\p f}{\p \ye}\nx \left.\frac{\p \ye}{\p \nb}\x \right) \nn \\
&+& dY_e\left(\left.\frac{\p f}{\p Y_e}\nye + \left.\frac{\p f}{\p \ye}\nx \left.\frac{\p \ye}{\p Y_e}\n \right)   \nn
\eeq
leading to
\beq
\left.\frac{\p f_2}{\p \nb}\x &=& \left.\frac{\p f}{\p \nb}\xye + (0)\left.\frac{\p \ye}{\p \nb}\x \nn \\
&=& \left.\frac{\p f}{\p \nb}\xye = \left.\frac{\p f_1}{\p \nb}\xye   \nn \\
\left.\frac{\p f_2}{\p Y_e}\n &=& \left.\frac{\p f}{\p Y_e}\nye + (0)\left.\frac{\p \ye}{\p Y_e}\n \nn \\
&=& \left.\frac{\p f}{\p Y_e}\nye = \left.\frac{\p f_1}{\p Y_e}\nye   ~. \nn
\eeq
This means that state functions such as the internal energy, the pressure, the entropy, etc.\ that depend of first derivatives of the free energy are impervious to the order in which equilibrium conditions are imposed and derivatives are performed. One might naively expect that second- and higher-order derivatives are similarly protected but that is not the case. In fact, the difference between the equilibrium and the adiabatic sound speeds arises precisely because the ordering is crucial ~\cite{Jaikumar:2021}.
\indent 
In what follows, we show results for the adiabatic sound speed in both the fast and the slow propagation limits which, we assume,  envelope physical reality.

\section{Results 
\label{sec:results}}

In this section, we discuss the impact of the local-to-total lepton ratio ($\eta = 0, \,0.1, \,0.3, \,0.6, \,1$) on the EOS and on matter composition. All quantities are presented for two different values of the net electron fraction ($Y_e = 0.1, \,0.4$) and of the temperature ($T=10,\,50$ MeV). 

%


\begin{figure}
    \centering  \includegraphics[width=.5\textwidth]{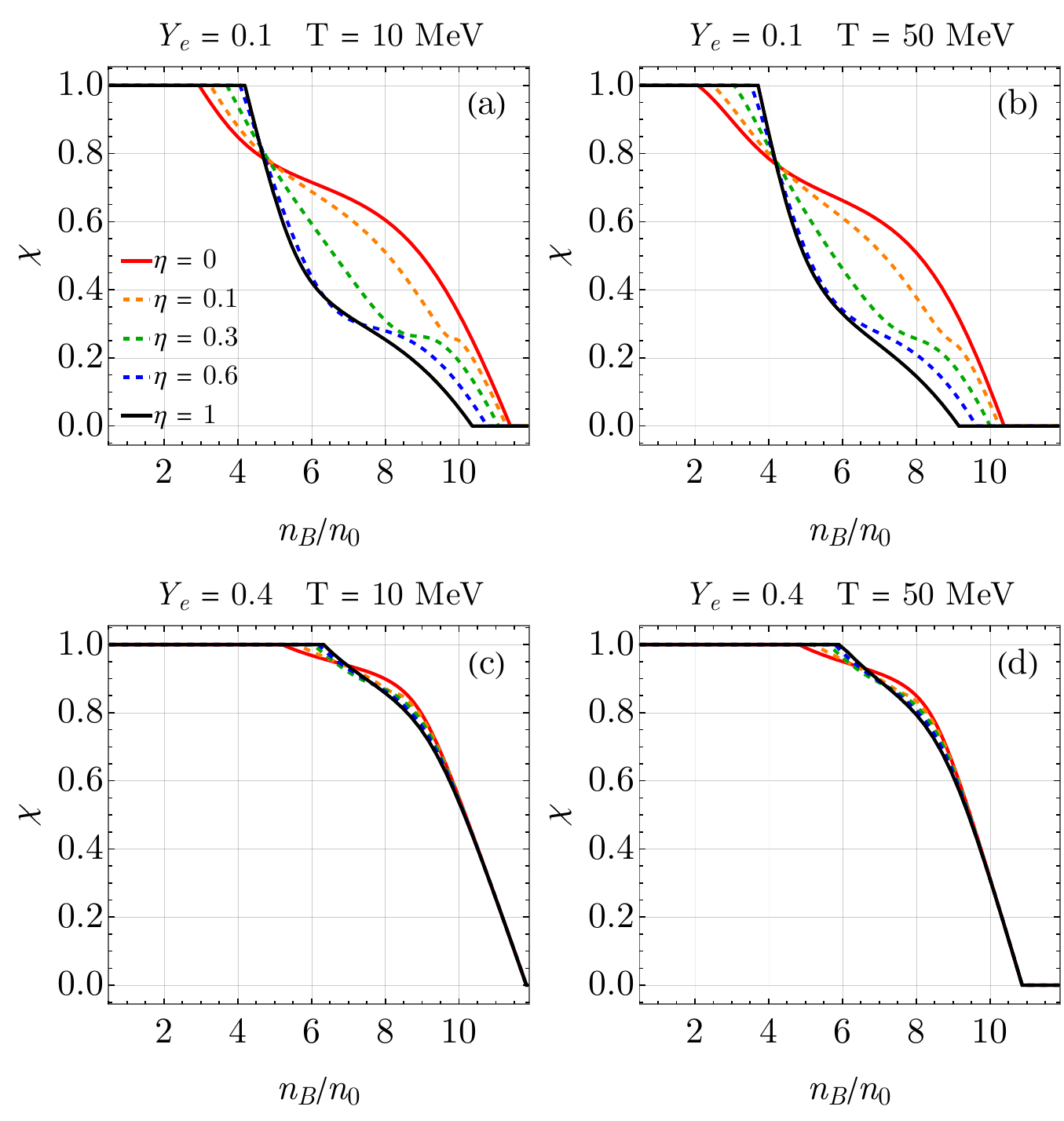} 
\caption{
The volume fraction of nucleons $\chi$ vs. baryon density $\nb$ in units of nuclear saturation density $n_0$ for the indicated values of the local-to-total lepton ratio $\eta$, 
with 
net electron fraction $Y_e=0.1$ and temperature $T=10$ MeV [panel (a)], 
$Y_e=0.1$ and $T=50$ MeV [panel (b)], 
$Y_e=0.4$ and $T=10$ MeV [panel (c)], 
and 
$Y_e=0.4$ and $T=50$ MeV [panel (d)]. 
\label{Fig:chi}}
\end{figure}

\begin{figure}
    \centering  \includegraphics[width=.5\textwidth]{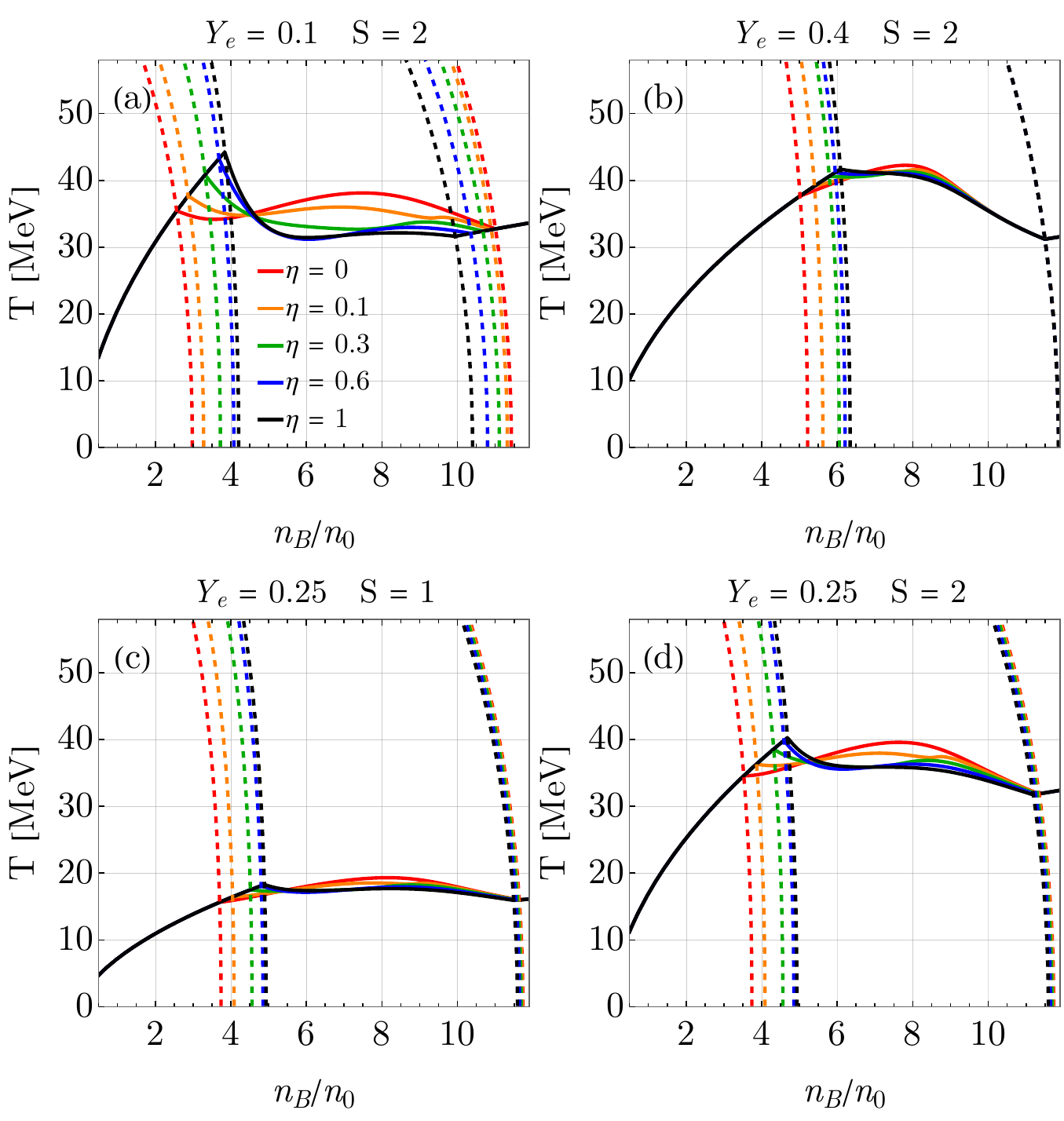} 
\caption{Phase diagram in the $T-\nb$ plane for the indicated values of the local-to-total lepton ratio $\eta$. 
For each case, phase boundaries($T$ vs. $\nb$) are shown as dashed(solid) lines at a fixed specific entropy $S=s/\nb$ (at either $S=1$ or $S=2$) and fixed $Y_e$ (either $Y_e=0.1$ or $Y_e=0.4$) as stated in the plot titles.
$Y_e=0.25$ cases [panels (c) and (d)] are reported here as well since Ref.~\cite{Kuroda:2021eiv} obtains it as a typical central value after the core bounce in a CCSN simulation of massive progenitors. \label{Fig:nTTS}}
\end{figure}

Figure~\ref{Fig:chi} shows the volume fraction of nucleons $\chi$ as a function of the baryon density $\nb$, while the phase diagram in the $T-\nb$ plane is reported in Fig.~\ref{Fig:nTTS}. The larger the local-to-total lepton ratio (i.e. the larger the surface tension), the narrower the mixed phase is in terms of baryon densities. As the temperature increases and the electron fraction decreases, quarks appear at lower baryon densities. 

\begin{figure}
    \centering  \includegraphics[width=.5\textwidth]{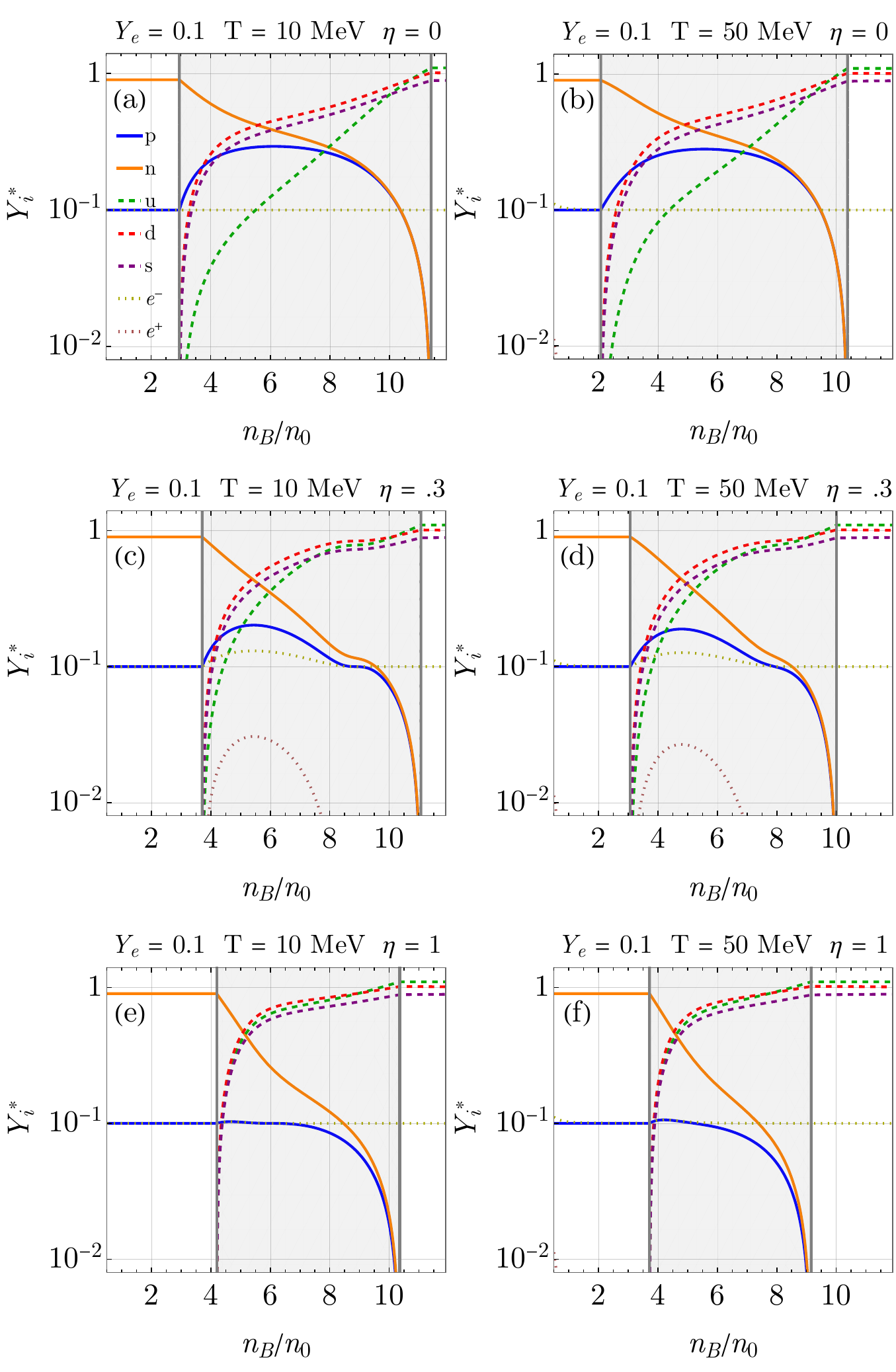} 
 \caption{Physical particle fractions $Y_i^*$ vs. baryon density $\nb$ at net electron fraction $Y_e=0.1$, for the indicated values of the local-to-total lepton ratio $\eta$ and temperature $T$. 
 Contributions from electrons and positrons are reported separately, while in the other cases, the net values are presented. 
\label{Fig:Y01}}
\end{figure}

\begin{figure}
    \centering  \includegraphics[width=.5\textwidth]{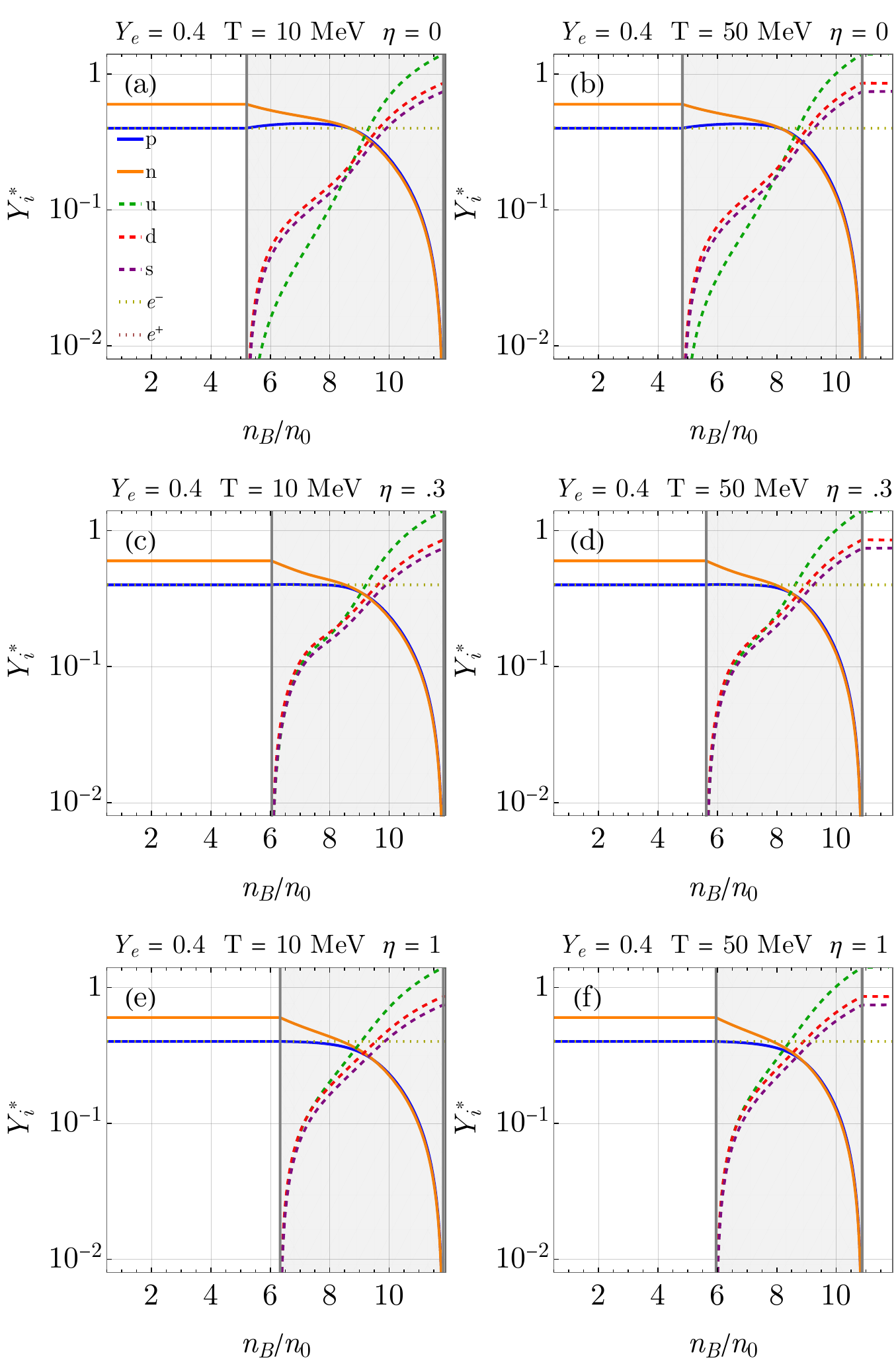} 
\caption{Same as Fig.~\ref{Fig:Y01} but at net electron fraction $Y_e=0.4$.
\label{Fig:Y04}}
\end{figure}

While the trend is fairly straightforward as the temperature increases, it needs a more precise discussion as the electronic fraction changes. Lower electron fractions imply lower proton fractions in the pure nucleonic phase, leading to a large symmetry energy. Thus, with a small electron fraction, the free energy of the nucleonic phase is higher, and the baryon density at which the presence of quarks minimizes the system's free energy becomes lower.
All the $\chi$ curves for different values of $\eta$ at fixed $Y_e, \,T$ have an approximate intersection where the free energy density of the pure nucleonic and quark phases are equal. Another relevant point is that, at high $Y_e$, the impact of the parameter $\eta$ becomes much less relevant in the early part of the mixed phase, while in the latter part, the curves and the end baryon density point of the mixed phase become almost $\eta$ independent. This behavior, also evident in all thermodynamic quantities below, occurs because, in nucleonic matter, the corresponding lepton fraction $Y_{\eN}$ can (at most) only slightly exceed 0.5. Thus, when the total lepton fraction $Y_e$ approaches 0.5, it becomes necessary for the lepton fraction associated with quarks $Y_{\eQ}$ to also be around 0.5 in order to satisfy lepton number conservation; that is, in the limit $Y_e \rightarrow 0.5$, the freedom to rearrange charges between the two phases is lost. Since, in this limit, all lepton fractions must have (approximately) the same value regardless of $\eta$, the distinction between the three types of leptons based on baryonic matter association (hadrons, quarks, both) becomes a moot point, and all $\eta$'s collapse onto the Gibbs case (single lepton cloud), $\eta=0$ (see also later discussion pertaining to Figs.~\ref{Fig:Y01} and~\ref{Fig:Y04}). \\

Note that the baryon number density and temperature of the mixed phase boundaries are parameter dependent. In particular, the bag parameter chosen in this work (as listed in Table~\ref{tab:pars}) leads to a mixed-phase onset density typically at $\nb\sim (8.3-8.4)\, n_0$ for symmetric matter, depending on $\eta$, in accordance with the constraints from heavy-ion experiments \cite{Danielewicz:2002pu}. 
This choice has been made in order to ensure a reasonably large extension in the range of $\nb$ for the mixed phase region, leading to a clearer, more straightforward interpretation of the qualitative behavior of thermodynamic quantities within it. Nevertheless, different choices of $B$ can be made to tune the density range that the mixed phase occupies, making it more in line with the ones used in astrophysical simulations such as \cite{Fischer:2017lag,Kuroda:2021eiv}. 

Figure~\ref{Fig:nTTS} illustrates the temperature $T$ as a function of the baryon density $\nb$ at fixed specific entropy $S$, providing an approximate evolutionary path for the central compact object after core bounce during a CCSN. In Ref.~\cite{Kuroda:2021eiv} terms, panel (c), with $S=1$, corresponds to a progenitor star of approximately 50 $\Msolar$, while panel (d), having $S=2$, is consistent with a progenitor around 80 $\Msolar$. CCSN simulations incorporating a quark-hadron phase transition under the Maxwell construction predict that a second shock may be triggered when the central density exceeds the quark-hadron transition density~\cite{Fischer:2017lag, Kuroda:2021eiv}. This secondary shock has the potential to drive a successful explosion. Analogously, BNSM simulations with similar EOSs featuring a first-order phase transition suggest that the post-merger GW signal may exhibit distinct signatures, such as a shift in the waveform phase~\cite{Most:2018eaw} or a shift in the peak frequency~\cite{Bauswein:2018bma}.
However, within our framework for phase transitions between the Maxwell and Gibbs constructions, the pressure and density variations are in general more gradual compared to the Maxwell scenario.
This smoother behavior may suppress the secondary shock in CCSNe and reduce the modification of the post-merger gravitational wave signal. In this regime, the first-order phase transition effectively resembles that of a crossover transition. Recent simulations of BNSMs involving hybrid stars with a crossover transition have shown that the post-merger peak frequency can be lower than in mergers involving either purely hadronic stars or hybrid stars undergoing a sharp first-order transition, provided the mass and tidal deformability are held fixed~\cite{Hensh:2024onv}. On the other hand, if the transition density is sufficiently high so that the phase transition occurs only in the post-merger remnant, with no influence during inspiral, the peak frequency may remain largely unaffected~\cite{Fujimoto:2024ymt}. Moreover, BNSMs involving a first-order phase transition under the Gibbs construction can be difficult to distinguish from purely hadronic mergers without a phase transition\cite{Prakash:2021wpz,Prakash:2023afe}.  Further numerical simulations properly incorporating the quark-hadron interface beyond the Maxwell or Gibbs construction are necessary to validate these findings and assess their implications. 

\begin{figure}
    \centering  \includegraphics[width=.5\textwidth]{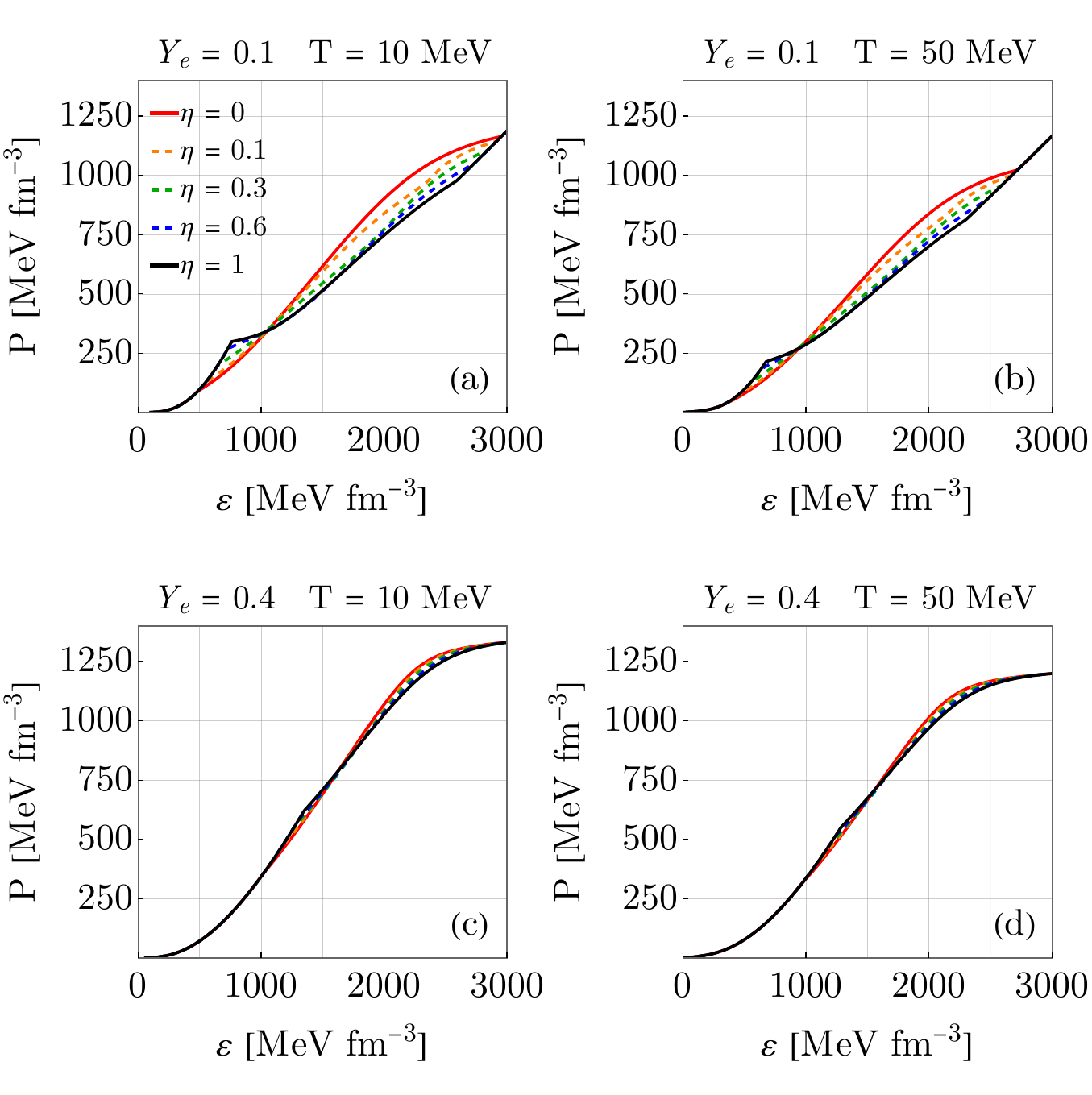}
\caption{Pressure $P$ vs. energy density $\ep$ for the indicated values of the local-to-total lepton ratio $\eta$, net electron fraction $Y_e$, and temperature $T$.  
Different from the $\beta$-equilibrium case, the pressure is not constant in the mixed phase even in the $\eta=1$ (Maxwell) case. 
\label{Fig:Pe}}
\end{figure}

Figure~\ref{Fig:Pe} shows the pressure $P$ as a function of the energy density $\ep$. As expected, the appearance of new degrees of freedom leads to a softening in the EOS (i.e. lower pressure at a fixed energy density). The general qualitative behavior as a function of the free variables $(Y_e,T,\eta)$ aligns with the previously described trends for $\chi$. \\

\begin{figure}
    \centering  \includegraphics[width=.5\textwidth]{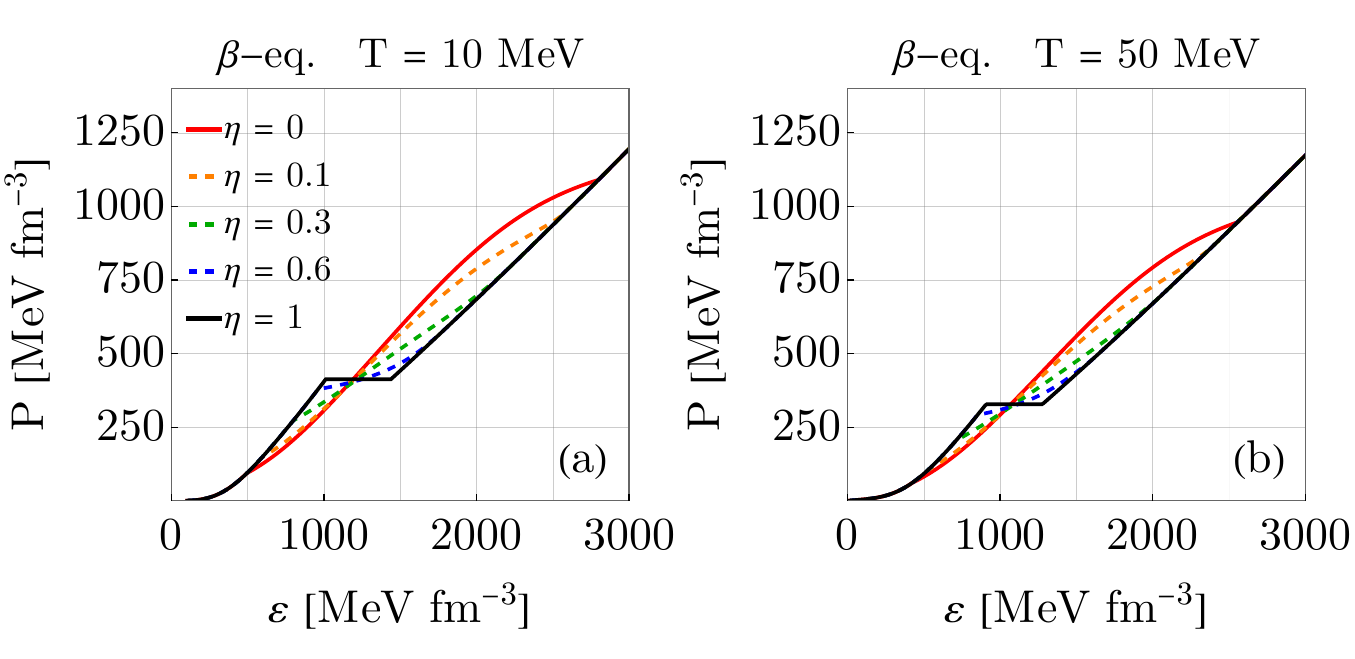} 
 \caption{Pressure $P$ vs. energy density $\ep$ in $\beta$-equilibrium for the indicated values of the local-to-total lepton ratio $\eta$, and temperature $T$.  \label{Fig:Pebeta}}
\end{figure}

Note that while in the $\beta$-equilibrium case reported in~\cite{Constantinou:2023ged} (see also Fig.~\ref{Fig:Pebeta}), the pressure is constant in the mixed phase for $\eta=1$ (i.e. MC) if $Y_e$ is an independent variable of the system, as in our case, the pressure is density dependent even in the $\eta=1$ mixed phase. In other words, in the context of first-order phase transitions, a construction that imposes a constant pressure as a function of energy density (or baryon density) does \textit{not} minimize the free energy of the system (i.e. is not a physical equilibrium solution), if the net electronic fraction is an independent variable of the system even in the case of fully local charge neutrality. \\

This behavior is due to the number of globally conserved charges in the system. It is, in fact, well known that a MC minimizes the free energy of the mixed phase if there is only one globally conserved charge~\cite{Glendenning:1992vb,Hempel:2009vp}. For example, in $\beta$-equilibrium and assuming fully local charge neutrality ($\eta=1$), the baryon number is the only globally conserved charge, and the pressure in the mixed phase is then flat.
In the absence of $\beta$ equilibrium in our approach, we always have at least two globally conserved charges: the baryon number [\Eqn{eqn:baryoncons}] and the non-leptonic electric charge $\chi Y_p+ (1-\chi) \left(2/3 \,Y_u-1/3 \,Y_d - 1/3 \,Y_s\right)$ [as can be noted by substituting Eqs.~(\ref{eqn:chargeneutH},~\ref{eqn:chargeneutQ},~\ref{eqn:chargeneutglob}) in \Eqn{eqn:leptoncons}] or, equivalently, the isospin. 

When $\eta=1$, electric charge neutrality is accomplished locally, and we are thus left with only two globally conserved charges (see case IIIb in~\cite{Hempel:2009vp}). However, when $\eta=0$, electric charge neutrality is globally achieved, and we therefore have a third globally conserved charge (see case V in~\cite{Hempel:2009vp}, but without neutrinos). \\

In a first-order phase transition, the two phases have in common one chemical potential for each globally conserved quantity. If $Y_e$ is an independent variable, the chemical potentials in common between the nucleonic and quark phases are at least two: \Eqn{eqn:strong1} and \Eqn{eqn:strong2}. Those chemical potentials can be interpreted as being related to ``strong neutral" and ``strong charged" charges. Note that the former is equivalent to the baryon chemical potential, while the latter can also be rewritten as the non-leptonic electric chemical potential ($\mu_p-\mu_n+\eta \mu_{\eN}=\mu_u-\mu_d+\eta \mu_{\eQ}$), namely the chemical potential associated with the non-leptonic electric charge $\chi Y_p+ (1-\chi) \left(2/3 \,Y_u-1/3 \,Y_d - 1/3 \,Y_s\right)$ considering charge neutrality as a constraint.
If $\eta=0$, the global charge conservation leads to a third common chemical potential, namely $\mu_{\eG}$.
In contrast, in $\beta$-equilibrium and $\eta=1$ (MC), the two phases have only one chemical potential in common, the baryon chemical potential. 
Indeed, \Eqn{eqn:strong2} 
can be rewritten as the $\beta$-equilibrium condition in the pure nucleonic phase ($\mu_p+\eta\mu_{\eN}+(1-\eta) \mu_{\eG}=\mu_n$) using \Eqn{eqn:strong1} and \Eqn{eqn:beta}, leaving 
\Eqn{eqn:strong1} 
as the only common chemical potential. 
If $\beta$-equilibrium is not assumed, then 
\Eqn{eqn:beta} does not apply and, thus, \Eqn{eqn:strong2} 
cannot be rewritten as a condition in the nucleonic phase alone, therefore both chemical potentials related to globally conserved charges remain.
Finally, in $\beta$-equilibrium and $\eta=0$ (GC),  global charge neutrality leads to a second common chemical potential, namely $\mu_{\eG}$.\\

Figures~\ref{Fig:Y01} and \ref{Fig:Y04} show
particle fractions $Y_i^*$ as  functions of the baryon density $\nb$. Note that contributions from electrons and positrons are reported separately, while curves for all other particles are referring to net fractions. 
As pointed out in~\cite{Constantinou:2023ged}, an advantage of using this approach is that we can maintain control over the various particle fractions even in the $\eta=1$ case. \\

The particle fractions shown in Fig.~\ref{Fig:Y01} and Fig.~\ref{Fig:Y04} exhibit minimal qualitative differences between 
$T=10$ MeV and $T=50$ MeV, except for the shift of the mixed region to lower densities with increasing temperature. Variations in particle fractions across different $Y_e$ and $\eta$ cases are primarily driven by the symmetry energy of the nucleonic sector and the strong equilibrium between nucleons and quarks. In the $\eta=0$ mixed phase, where $Y_{\eG}^*=Y_e$, the electron chemical potential $\mu_{\eG}$ increases with density following a free Fermi gas behavior ($\propto \nb^{1/3}$). 
The positive $\mu_{n}-\mu_{p}=\mu_{d}-\mu_{u}$ leads to enhanced populations of the negatively charged $d$ and $s$ quarks relative to the positively charged $u$ quark at lower densities in the mixed phase. To balance with the negatively charged quark matter, proton fraction $Y_p^*$ increases rapidly until $Y_p^*\approx Y_n^*$, at which $\mu_{n}-\mu_{p}\ll \mu_{n}+\mu_{p}$. 

In the $\eta=1$ mixed phase with the total $Y_e$ fixed, the local components $Y_{eN}^*$ and $Y_{eQ}^*$ are not fixed. Initially, we note a smaller increase in $Y_{eN}^*$, which is related to a decrease in $Y_{eQ}^*$ due to an excess of positrons in the local quark sector. As density increases, $Y_{eQ}^*$ remains small due to the presence of negatively charged $d$ and $s$ quarks, leading to $Y_p^*=Y_{eN}^*\approx Y_e$. As the nucleonic sector becomes less neutron-rich with increasing density, strong equilibrium forces the quark sector to become more positively charged, causing an increase in $Y_{eQ}$. Taking volume fraction into account, the relation $Y_{eQ}^*=(1-\chi) Y_{eQ}$ further enhances this increase, allowing $Y_{eN}^*$ to decrease.

For intermediate $\eta$, the qualitative behavior of baryonic fractions lies between the two extreme cases. The initial increase of $Y_p^*$ is balanced partially by the global $Y_{eG}^*$, and partially by the increase of $Y_{eN}^*$  counteracted by a positron excess in the quark sector. Thus, the main qualitative difference between $\eta$ cases is that at small $\eta$, $Y_p^*$ increases in the early part of the mixed phase, while at high $\eta$, it remains nearly constant.

At large $Y_e$, in the pure nucleonic phase, the contribution of symmetry energy to the nucleon chemical potential is lower, delaying the onset of quarks. As in the previous discussion, $Y_p^*$ increases with density immediately after the appearance of quarks. However, the magnitude of this increase is significantly reduced, as the pure hadronic phase is no longer neutron-rich at large $Y_e$. As density increases further, the hadronic sector becomes even less neutron-rich, quickly reaching $Y_p^*=Y_n^*$ and being constrained from becoming proton-rich due to the symmetry energy, maintaining $Y_p^*\approx Y_n^*$.

We note that at large $Y_e\sim 0.5$, particle fractions and thermodynamic behavior are nearly $\eta$-independent. Generally speaking, the quark phase has a stronger tendency to keep small $Y_{eQ}$ due to the presence of negatively-charged quarks. In the mixed phase for physically relevant values $Y_e<0.5$, we observe $Y_{eN}>Y_e>Y_{eQ}$. The smaller $\eta$, the higher $Y_{eN}$ and the smaller $Y_{eQ}$. However, the positive symmetry energy constrains $Y_{eN}$ to be smaller than about 0.5. Thus, for $Y_e \sim 0.5$, the computed $Y_{eN}$ remains around $\sim 0.5$ at different values of $\eta$ in the mixed region, while $Y_{p}^*=\chi Y_{p}=\chi Y_{eN}$ decreases with $\chi$. This behavior is general, but the large symmetry energy of ZL EOS at high baryon density makes it more prominent.

\begin{figure}
    \centering  \includegraphics[width=.5\textwidth]{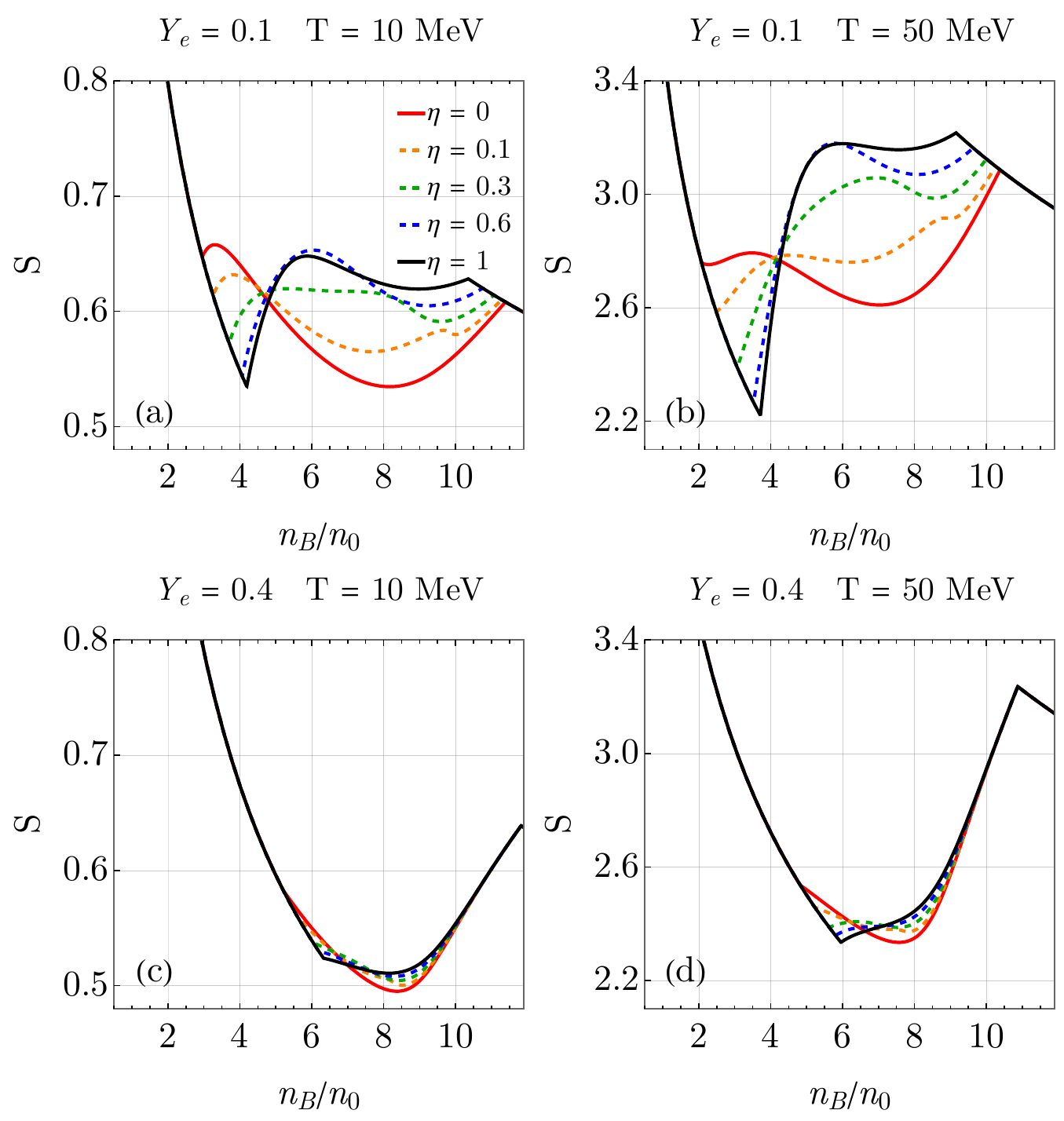} 
\caption{Specific entropy $S=s/\nb$ vs. baryon density $\nb$ for the indicated values of the local-to-total lepton ratio $\eta$, net electron fraction $Y_e$, and temperature $T$. 
\label{Fig:S}}
\end{figure}

\begin{figure}
    \centering  \includegraphics[width=.5\textwidth]{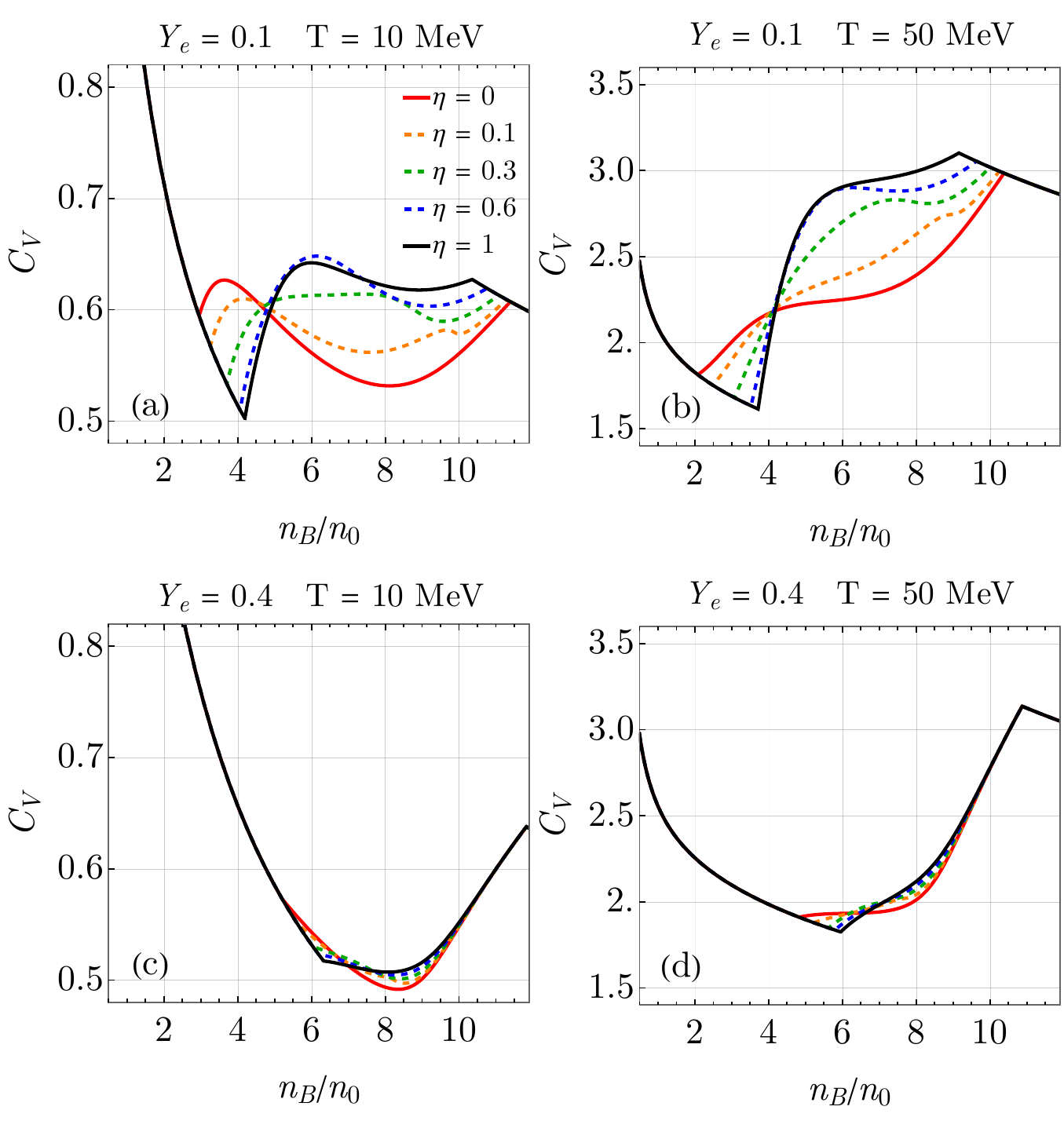} 
\caption{Specific heat per baryon at constant volume $C_V$ vs. baryon density $\nb$ for the indicated values of the local-to-total lepton ratio $\eta$, net electron fraction $Y_e$, and temperature $T$. 
\label{Fig:Cv}}
\end{figure}

\begin{figure}
    \centering  \includegraphics[width=.5\textwidth]{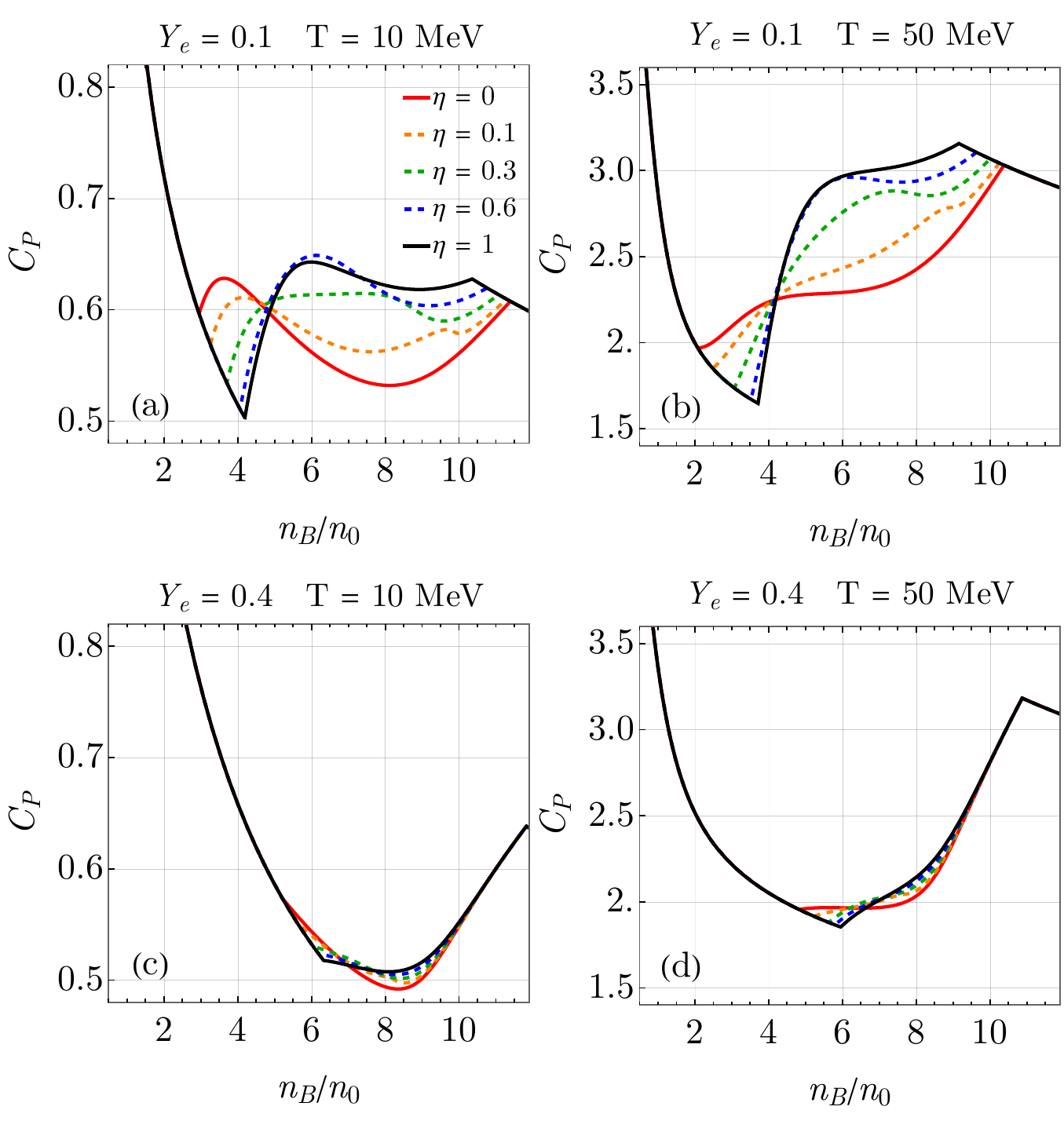}
\caption{Specific heat per baryon at constant pressure $C_P$ vs. baryon density $\nb$ for the indicated values of the local-to-total lepton ratio $\eta$, net electron fraction $Y_e$, and temperature $T$. 
\label{Fig:Cp}}
\end{figure}

The appearance of quark degrees of freedom leads to an increase in the specific entropy $S=s/\nb$ with respect to the pure nucleonic case as shown in Fig.~\ref{Fig:S}. 
The specific entropy decreases monotonically as a function of the baryon density in the pure phases, while in the mixed phase it is nonmonotonic. Its qualitative behavior, as determined by $\eta$,  follows a trend similar to that seen in other thermodynamic quantities. 
The specific entropy also increases monotonically as a function of the temperature. Moreover, the difference between $S=s/\nb$ in the pure quark and pure nucleon phases is higher with increasing temperature. 
The same qualitative behavior can be found in the heat capacity at constant volume $C_V$ and at constant pressure $C_P$ reported in Figs. \ref{Fig:Cv} and \ref{Fig:Cp} respectively. Note that the two heat capacities differ mainly in the low-density regime.

We point out that, in the presence of a phase transition, the adiabatic compressibility $\partial P/\partial \nb|_S$ can become negative which implies a mechanical instability. Consequently, related response functions such as the specific heat at constant pressure $C_P$ [\Eqn{eqn:cpnt}] may exhibit divergences or discontinuities. Here, different models are invoked for the description of the two phases. The switch from one model to the other is accompanied by a negative compressibility, being that the quark phase is energetically favored above the transition density leading to a drop in the pressure. 

It is this unphysical behavior of the compressibility that the Maxwell and Gibbs constructions (and, indeed, our generalization thereof) address at a macroscopic level. The enforcement of mechanical equilibrium via \Eqn{eqn:meceq} ensures that the pressure never decreases over the phase coexistence region. As discussed in Sec.~\ref{sec:adss} this condition is imposed after performing partial derivatives (meaning that the latter are taken at a fixed $\chi$) guaranteeing their continuity; thus $\partial P/\partial \nb|_S > 0$ always. Accordingly, any features/extrema in $C_P$ are indications of the mixed phase but \textit{not} of an instability. This is still true for constructions of matter in $\beta$ equilibrium with $\eta \sim 1$ which have a constant pressure over the mixed phase. Thus, even equations of state of this kind of matter generate perfectly stable compact stars, albeit ones with vanishingly small regions/amounts of hybrid matter and attendant jumps in $\nb$ (from a hadronic shell to a quark core) as a function of the radius of the star.

In hydrodynamic simulations, conserved quantities such as the total energy density and the baryon density are typically used as base variables, as they satisfy the continuity equations. Other thermodynamic quantities, such as temperature and pressure, are then derived as needed from these base variables. Total pressure can be decomposed into the zero-temperature pressure and the thermal pressure,
\begin{eqnarray}
P(\ep,\nb)&=&P(\nb,T=0)+P_{th}(\nb,T) \label{eq:Pth}\\
\Gamma(\nb,T)&=&\frac{P_{th}(\nb,T)}{\ep - \ep(\nb,T=0)}+1 \label{eq:Gamma}
\end{eqnarray}
where the thermal index $\Gamma(\nb,T)$  can be treated as a constant known as the thermal index or $\Gamma$-law approximation. 
As a result, knowing the cold EOS, $P(\nb,T=0)$ and $\ep(\nb,T=0)$, plus the thermal index $\Gamma$ is sufficient rather than taking a 2-D interpolation of $P(\ep,\nb)$. This approximation works extremely well in the case of relativistic ($\Gamma=4/3$) and non-relativistic ($\Gamma=5/3$) single-component degenerate Fermi gas where the temperature dependence of $\Gamma$ is weak. 
It is also possible to use a density-dependent (instead of a constant) $\Gamma$ to account for the equilibrium of multicomponent matter as another 1-D interpolation $\Gamma(\nb,T=0)$ similar to $P(\nb,T=0)$ and $\ep(\nb,T=0)$. 

\begin{figure}
    \centering  \includegraphics[width=.5\textwidth]{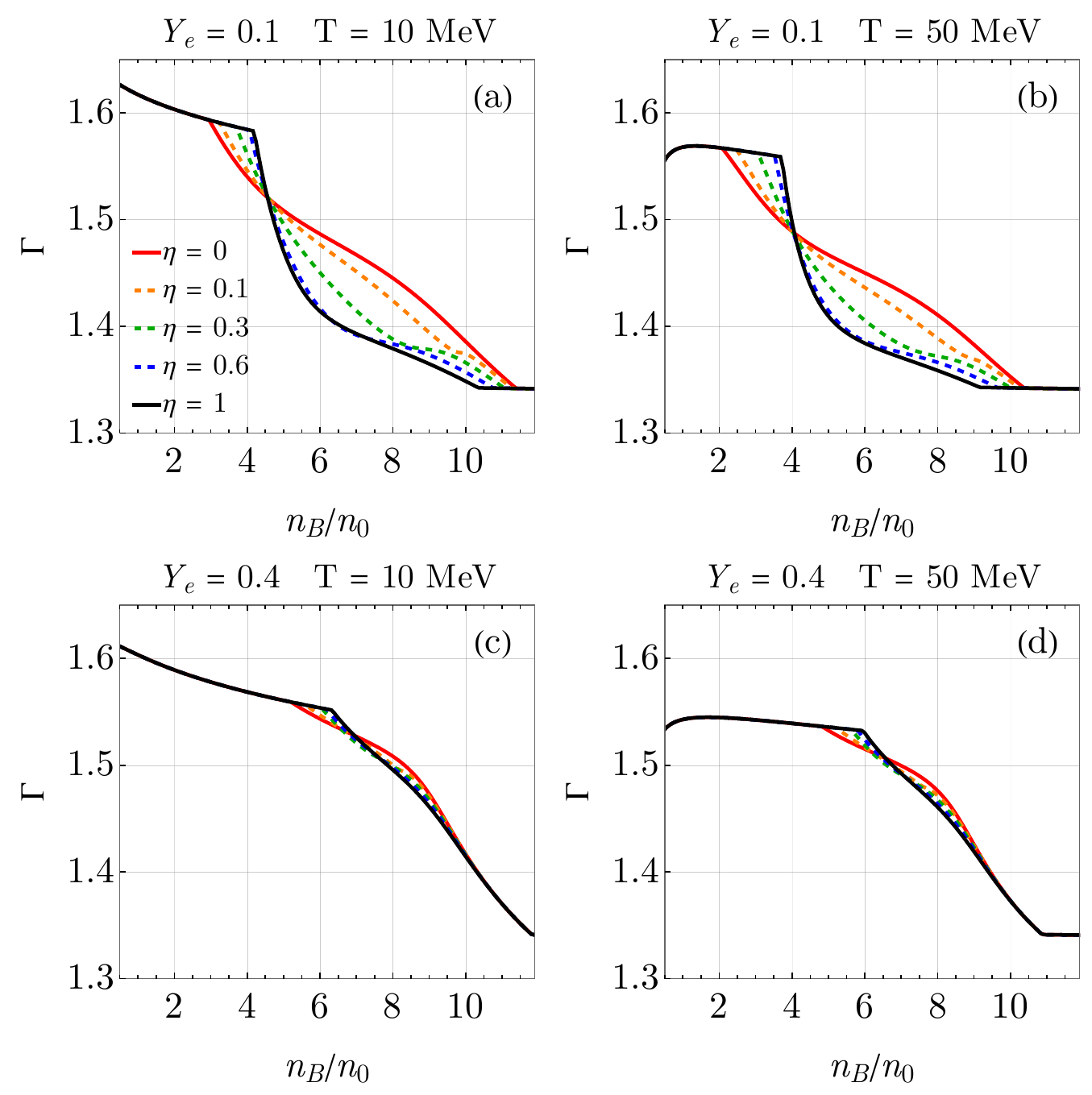} 
\caption{
Thermal index using definition (ii) vs. baryon density $\nb$ for the indicated values of the local-to-total lepton ratio $\eta$, net electron fraction $Y_e$, and temperature $T$. 
}
\label{Fig:gamma_fix_chi}
\end{figure}

For EOSs in CCSN simulations, the proton fraction $Y_e$ is also a free variable in \Eqn{eq:Pth} and \Eqn{eq:Gamma}, leading to $P(\nb,Y_e)$ and $\ep(\nb,Y_e)$ as the zero-temperature EOS. 
Fortunately, the dependence of $\Gamma(\nb,Y_e,T)$ on $Y_e$ is weak for both hadronic and quark matter, making the constant $\Gamma$ approximation still reasonable. 
This can be validated by comparing the upper [(a) and (b)] and lower [(c) and (d)]  panels of Fig.~\ref{Fig:gamma_fix_chi}. However, the situation becomes more intricate in the case of a hybrid EOS, where a volume fraction $\chi$ is introduced that is determined by mechanical and chemical equilibrium between phases. Therefore, thermal pressure may be defined in various ways: 
\\
\begin{enumerate}[(i)]
    \item  $P_{th}=P\left[{\{Y_i(T=0)\}},\chi{(T=0)}, T\right]-P\left[{\{Y_i(T=0)\}},\chi{(T=0)},T=0\right]$
    \item  $P_{th}=P\left[{\{Y_i(T)\}},\chi{(T)}, T\right]-P\left[{\{Y_i(T)\}},\chi{(T)},T=0\right]$
    \item  $P_{th}=P\left[{\{Y_i(T)\}},\chi{(T)}, T\right]-P\left[{\{Y_i(T=0)\}},\chi{(T=0)},T=0\right]$
\end{enumerate}
where the variables $\nb$ and $Y_e$ are omitted, namely $P(\{Y_i\},\chi,T)$ stands for $P(\nb,\{Y_i\},\chi, T)$, while $\{Y_i(T)\}$ and $\chi{(T)}$ stands for $\{Y_i(\nb,Y_e, T)\}$ and $\chi(\nb,Y_e, T)$, where the $(\nb,Y_e, T)$ dependence comes from equilibrium conditions reported in Sec.~\ref{sec:fwork}. 
Thermal energy density $\ep_{th}$ can be defined accordingly. 
For the calculation of (i), we begin with the $T=0$ pressure of a specific composition, then compute the finite-$T$ pressure of the identical composition, and finally take their difference. In (ii), thermal pressure is defined as the total pressure at a given temperature $T$ less the nonthermal (or, equivalently, the exclusively density-dependent) contributions of that particular configuration at the same temperature, whereas in (iii), it is given by the difference between the pressure at $T$ and the pressure at $T=0$. In the pure phases, the three definitions coincide but in the mixed phase, whose boundaries depend on the temperature, (i) and (ii) have markedly different behavior compared to (iii). In the latter case, only the net lepton fraction $Y_e$ is the same between $T=0$ and finite-$T$ and, in fact, there exist combinations of $\nb$ and $Y_e$ for which only one of the two temperatures corresponds to the mixed phase.

\begin{figure}
    \centering  \includegraphics[width=.5\textwidth]{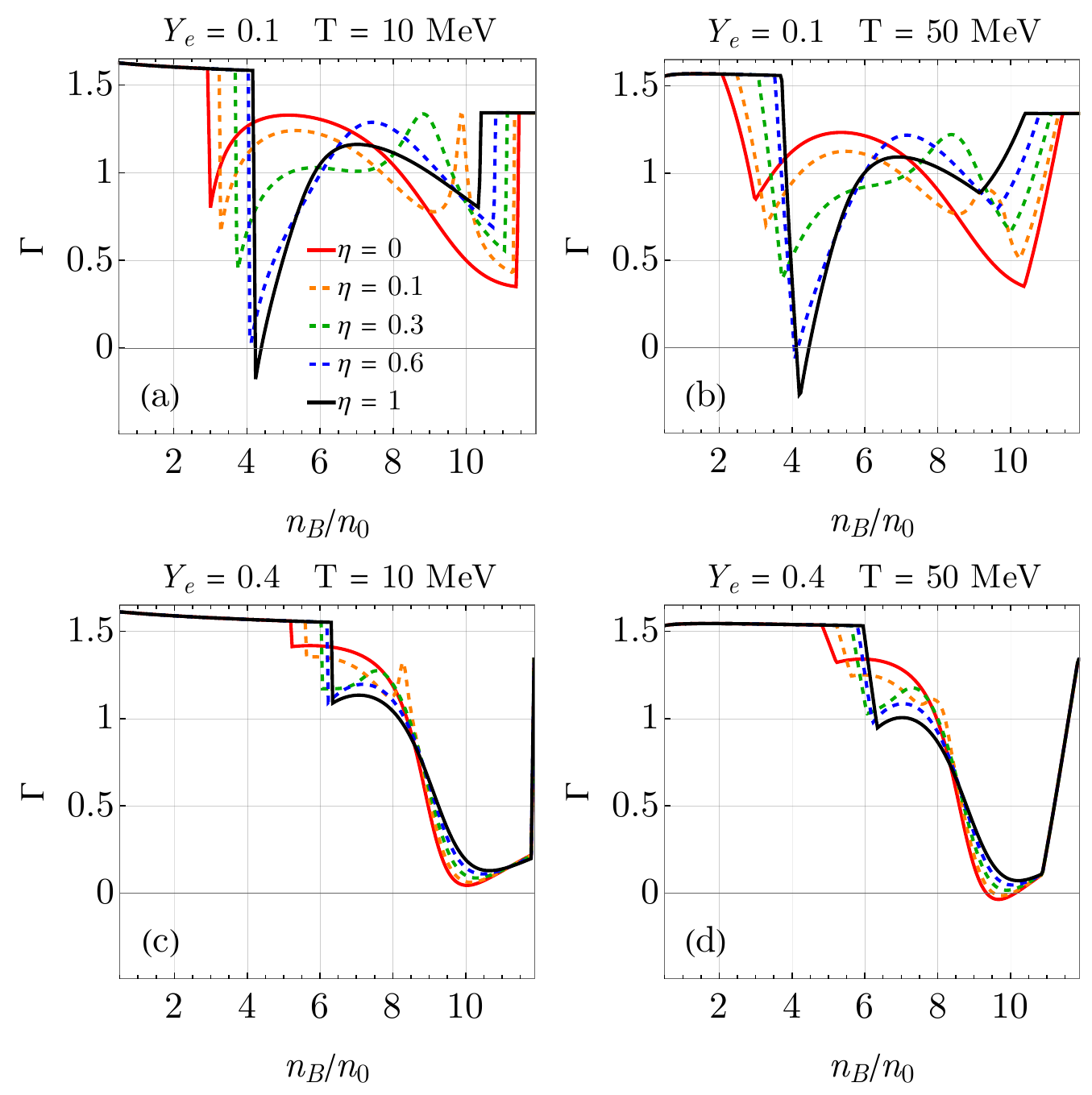} 
\caption{Thermal index using definition (iii) vs. baryon density $\nb$ for the indicated values of the local-to-total lepton ratio $\eta$, net electron fraction $Y_e$, and temperature $T$.
}
\label{Fig:gamma_vary_chi} 
\end{figure}

We compute the thermal index as $\Gamma=1+P_{th}/\ep_{th}$ for various cases and display the results in Fig.~\ref{Fig:gamma_fix_chi} and Fig.~\ref{Fig:gamma_vary_chi}. 
Definitions of the thermal pressure using (i) and (ii) are qualitatively the same, especially for low temperatures, therefore we only show definition (ii) in Fig.~\ref{Fig:gamma_fix_chi}. \\
In the mixed phase region, the thermal index monotonically decreases with density from about 5/3 for hadronic matter to 4/3 for quark matter. Varying charge neutrality from GCN ($\eta=0$) to LCN ($\eta=1$), the thermal index exhibits a steeper drop, due to the shrinking of the size of the mixed region.

Figure~\ref{Fig:gamma_vary_chi} shows the thermal index as defined in (iii). 
With the definition (iii), the thermal index behavior can be divided into five different regions in terms of baryon density (see also \cite{Blacker:2023afl,Blacker:2024wfg}). The first (last) one is the region of baryon densities at which the system is in the pure nucleonic (quark) phase at both finite $T$ and $T=0$. In that region, the thermal index is the one of pure nucleonic (quark) matter and is thus nearly constant. The second region corresponds to the baryon densities at which the system is in the mixed phase for finite $T$, but still in the pure nucleonic phase for $T=0$. There, the thermal index drops sharply and monotonically. Similarly, in the fourth region the system is in pure quark phase for finite $T$, but still in mixed phase for $T=0$, and the thermal index grows quickly monotonically. 
Finally, the third region corresponds to the baryon density range at which the system is in mixed phase for both finite $T$ and $T=0$.  The thermal index in this mixed region is non-monotonic and can have multiple peaks, with values significantly smaller than the nearly constant values attained in the nucleonic and quark phases. 

At both ends of the third region in the mixed phase, the thermal index is usually smaller than in the middle of the region and can even dip below zero for some specific range of $\nb$ and $Y_e$.

Note that regions two and four may be erroneously mistaken for discontinuity points.  Instead, they are small regions in which the thermal index decreases and increases rapidly, and whose size in terms of baryon density increases with increasing temperature. This effect, present using definition (iii), is due to the fact that the mixed phase boundaries are at different baryonic densities at different temperatures. The same effect is not present in the other definitions since, in them, $\chi$ is held constant in the calculations of thermal pressure and thermal energy density. Only in the limit $T\rightarrow 0$ the second and fourth regions collapse into two single points of baryon density, in which the thermal index is discontinuous. 

To understand this, we can take the limit of $T\rightarrow 0$, so that the three definitions of the thermal pressure reduce to (i) 
$P_{th}/T=\frac{\partial P}{\partial T}|_{\nb,Y_e,\{Y_i\},\chi}(T=0)$; 
(ii) 
$P_{th}/T=\frac{\partial P}{\partial T}|_{\nb,Y_e,\{Y_i\},\chi}(T)$; 
(iii) 
$P_{th}/T=\frac{\partial P}{\partial T}|_{\nb,Y_e}(T=0)$. 
There is no difference between (i) and (ii) since we already take $T\rightarrow 0$. The difference between (i) and (iii) can be understood as,
\begin{eqnarray}
\frac{\partial P}{\partial T}|_{\nb,Y_e} &=& \frac{\partial P}{\partial T}|_{\nb,Y_e,\{Y_i\},\chi}+\frac{\partial P}{\partial \{Y_i\}}|_{\nb,Y_e,\chi,T}\frac{\partial \{Y_i\}}{\partial T}|_{\nb,Y_e}\nonumber\\
&&+\frac{\partial P}{\partial \chi}|_{\nb,Y_e,\{Y_i\},T}\frac{\partial \chi}{\partial T}|_{\nb,Y_e}
\end{eqnarray}
where the third term on the right-hand side is zero for both pure hadronic phase ($\chi=1$) and pure quark phase ($\chi=0$), while taking a finite value in the mixed phase as shown in Fig.~\ref{Fig:chi}. 

Thus, using a constant thermal index to extend a cold EOS to finite temperatures is inadequate to describe the physics of the mixed phase. Furthermore, using an interpolation between two constant thermal indices of the pure nucleonic and quark phases for the mixed phase can only be a good choice if definitions (i) or (ii) are used, that is, if the extension to finite temperatures is done \textit{before} fixing the quantities $\{Y_i\},\chi\rightarrow \{Y_i(\nb,Y_e,T)\},\chi(\nb,Y_e,T)$ using the proper equilibrium conditions.
A method to approximate the thermal index using definition (iii) is proposed in \cite{Blacker:2023afl,Blacker:2024wfg}.

\begin{figure}
    \centering  \includegraphics[width=.5\textwidth]{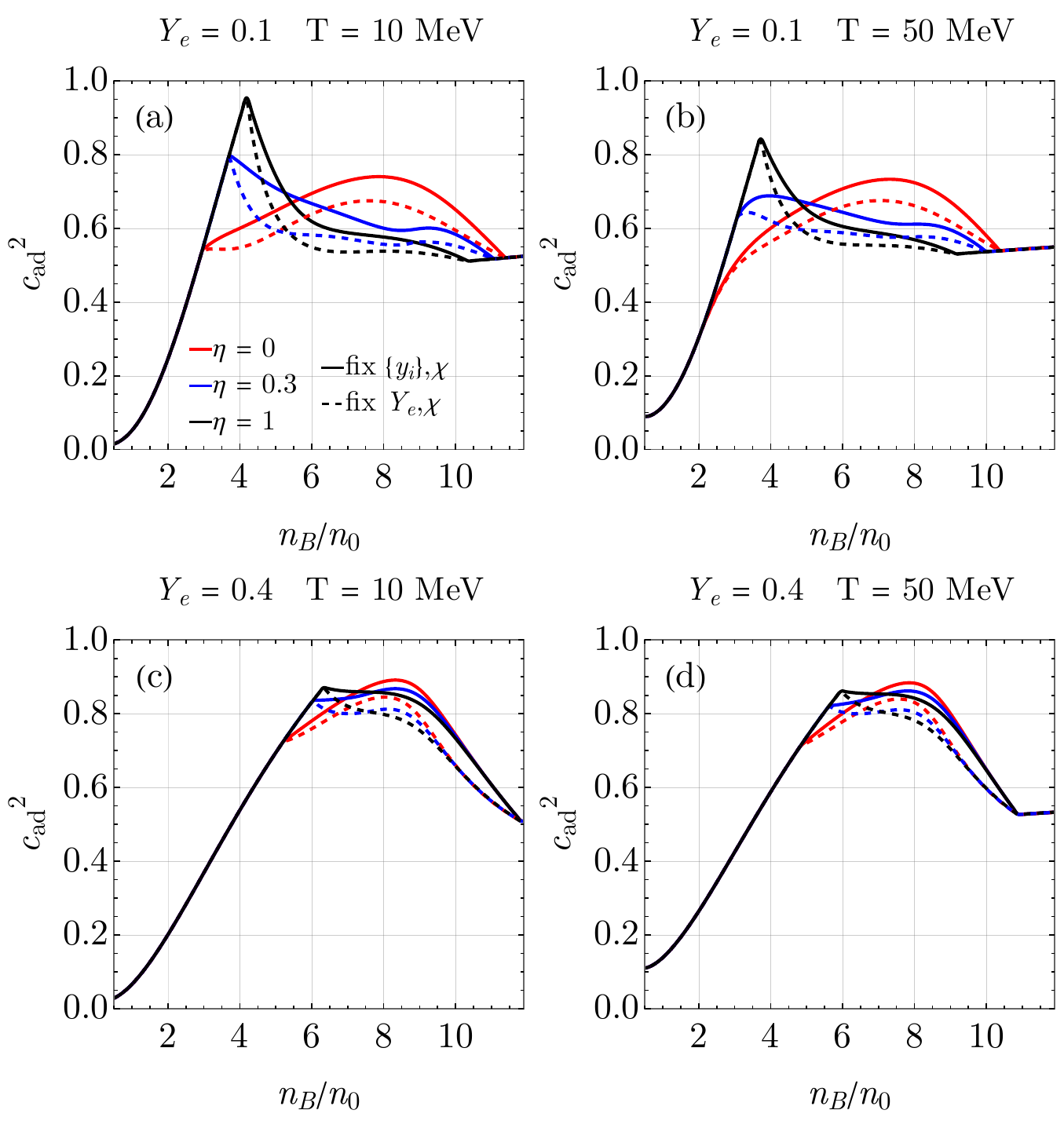} 
\caption{Adiabatic sound speed squared $\cad^2$ vs. baryon density $\nb$ in the fast-propagation (fixed $\{y_i\}, \chi$) and in the slow-propagation (fixed $Y_e, \chi$) regimes, for the indicated values of the local-to-total lepton ratio $\eta$, net electron fraction $Y_e$, and temperature $T$.} \label{Fig:cad2slowfastchifixed} 
\end{figure}

The adiabatic sound speed squared in the fast and slow fluctuation propagation (due to some external perturbation)
cases (see discussions in Sec.~\ref{sec:numer}) at fixed ($y_i, \chi$) and fixed ($Y_e, \chi$), respectively, are shown in Fig.~\ref{Fig:cad2slowfastchifixed}. 
In general, the sound speed reaches a peak and then decreases until it reaches the pure quark phase. For high values of $\eta$, the peak is reached at the density at which the mixed phase begins, after which the sound speed decreases monotonically for the entire mixed phase. For low values of $\eta$, on the other hand, the mixed phase starts with relatively lower values of sound speed and reaches a peak in the mixed phase. Moreover, we note that in the slow-propagation regime, the sound speed in the mixed phase is, in general, slightly lower than in the fast-propagation case.
We also point out that the squared speed of sound is approximately 1/2 in the quark phase, exceeding the conformal limit of 1/3. This increase is due to the repulsive vector interaction, introduced in the vMIT bag model to emulate the non-perturbative effects expected at these densities, well below the asymptotic pQCD regime.

Finally, we report for comparison the pressure as a function of the energy density (Fig.~\ref{Fig:Pebeta}) in neutrino-less $\beta$-equilibrium, namely, $Y_e$ is not an independent variable anymore but is computed for each $\nb$ and $T$ using \Eqn{eqn:beta}. 
As pointed out before, imposing neutrinoless $\beta$-equilibrium and fully local charge neutrality $\eta=1$, the system has only one globally conserved charge, and the iso-thermal pressure is then constant in the mixed phase as a function of the energy density and baryon density. 
However, one should consider that neutrinos-less $\beta$-equilibrium ($\mu_n=\mu_p+\mu_e$) is exact only at $T=0$~\cite{Alford:2018lhf,Alford:2021ogv}. 
At high temperatures, the neutrino mean free path becomes smaller than the size of the system, and neutrinos become trapped and in equilibrium with the rest of the matter. This possibility can be explored in our framework by adding the lepton fraction $Y_L=Y_e+Y_{\nu_e}$ as an independent variable, the equilibrium condition $\mu_{\nu_eH}=\mu_{\nu_eQ}$ related to the global conservation of the lepton number and replacing \Eqn{eqn:beta} with $\mu_d+\mu_{\nu_eQ} = \mu_u + \eta \mu_{\eQ} + (1-\eta)\mu_{\eG}  $. Moreover, in the intermediate conditions in which matter is too hot for neutrinos to be completely ignored but too cold for neutrinos to be trapped and in equilibrium with matter, a proper treatment would need calculations of reaction rates. In particular, Refs.~\cite{Alford:2018lhf,Alford:2021ogv} show that under these intermediate conditions, electron capture is much less suppressed than neutron decay (i.e. detailed balance is not fulfilled), and the $\beta$-equilibrium condition used at $T=0$ necessitates a model-dependent correction. 

We stress that the main goal of this work is to provide a finite-temperature EOS with $Y_e$ left as an independent variable without considering neutrinos, whose contribution is to be properly added in simulations. 

\section{Summary and conclusions}
\label{sec:sum}

The main goal of this work was to extend the framework for hadron-to-quark first-order phase transitions described in~\cite{Constantinou:2023ged} to finite temperatures and out-of-$\beta$ equilibrium conditions. This generalization is crucial for applications in simulations of high-energy astrophysical phenomena such as CCSNe and BNSMs.

Our approach introduces a continuous parameter $\eta$ that allows for a controlled continuous spectrum of thermodynamically-consistent constructions between the extreme cases of Maxwell and Gibbs (corresponding to local and global charge neutrality), thereby capturing intermediate cases between LCN and GCN. 
Using local or global charge neutrality simulates, within a thermodynamical framework, Coulomb charge screening due to the long-range nature of Coulomb interaction. 
In particular, LCN and GCN are good descriptions when the sizes of the structures in the mixed phase (controlled by the surface tension) are respectively much bigger and much smaller than the electron's Debye screening length.

We have analyzed the impact of the parameter $\eta$ in different thermodynamic quantities, including the pressure, energy density, entropy,  specific heats, and speed of sound at different baryon densities, temperatures and net electron fractions. 
We have found that higher temperatures and lower net electron fractions lead to a decrease in the onset densities for the mixed and also the pure quark phases. While increasing $\eta$, the mixed phase becomes narrower in terms of baryon densities.

We have shown that, unlike in the $\beta$-equilibrium case —where the pressure remains constant in the mixed phase at finite temperatures— the pressure in the out-of-$\beta$-equilibrium case depends on density, temperature, and electron fraction $Y_e$. When $Y_e$ and temperature are fixed, the pressure is no longer constant in the mixed phase, even in the fully LCN case.
Indeed, the MC is the correct equilibrium solution for the mixed phase only when one charge is globally conserved. However, if $Y_e$ is constant, at least two globally conserved charges are present:  baryon number and isospin.

We noted that for matter with high $Y_e$, thermodynamic quantities become almost $\eta$ independent. This is because quark matter has a strong tendency to maintain a very low $Y_{\eQ}$ as it accommodates negatively charged $d$ quarks in place of electrons. Consequently, electrons are redistributed among the nucleonic and quark components in the mixed phase, leading to $Y_{\eQ}<Y_e<Y_{\eN}$ when $Y_e<0.5$. The discrepancy between $Y_{\eN}$ and $Y_e$ increases with $\eta$, because more electrons are associated with a particular nucleonic or quark component. However, the trend ends when $Y_{\eN}\lesssim 0.5$ since the symmetry energy in nucleonic matter disfavors $Y_{\eN}> 0.5$. Thus, if $Y_e \simeq 0.5$,  $Y_{\eN}$ cannot go much beyond 0.5, which limits the discrepancy between $Y_{\eN}$ and $Y_e$, making the various state and response functions largely independent of $\eta$.

We computed the thermal index $\Gamma$, which is sometimes employed to simply extend zero-temperature EOSs to the finite temperature regime, using two different approaches. We have shown that although a constant thermal index is a good approximation for a pure nucleonic EOS, the mixed phase of a first-order phase transition would need a different treatment. Moreover, we noted that the behavior of the thermal index in the mixed phase strongly depends on the manner in which thermal and cold contributions to the pressure and energy are separated.

In this work, we use nucleons as the only hadronic degrees of freedom, with an EOS based on an energy density functional. However,  this framework can be easily applied to other EOS models for hadrons and quarks. In particular, it would be interesting to study the effects of including hyperonic degrees of freedom in the hadronic sector and of a color-superconducting phase in the quark sector. In general, the presence of hyperons and deltas will soften the hadronic EOS, and will make it more energetically favorable (i.e. lower $\mathcal{F}$ vs $\nb$ and higher $P$ vs $\mub$ assuming $\beta$-equilibrium) after the onset of hyperons and deltas. This will lead to a shift of the mixed phase to higher baryon densities. On the other hand, color superconductivity can stiffen the deconfined quark phase and make it more energetically favorable, lowering the onset density of the mixed phase. In some cases, the mixed phase emerges at a lower baryon density than the one at which hyperons appear in purely hadronic matter \cite{Ivanytskyi:2022oxv,Gartlein:2023vif}. Moreover, color superconductivity can also lead to an onset density that decreases nonmonotonically with temperature, due to the temperature dependence of the gap.

In the present work, $\beta$-equilibrium is not assumed (except where explicitly stated, see Fig. \ref{Fig:Pebeta}) and neutrinos are not included. The inclusion of neutrinos can be easily accomplished by introducing a new independent variable $Y_L=Y_{\nu_e}+Y_e$ and imposing the condition $\mu_{\nu_eH}=\mu_{\nu_eQ}$ (assuming lepton number is globally conserved). Moreover, if trapped neutrinos are in equilibrium with the rest of matter, the electron fraction $Y_e$ can be fixed using the condition $\mu_d+\mu_{\nu_eQ} = \mu_u + \eta \mu_{\eQ} + (1-\eta)\mu_{\eG}  $. 
However, neutrinos cannot always be assumed to be in thermodynamic equilibrium in BNSMs and CCSNe \cite{CompOSECoreTeam:2022ddl}.
Modern numerical simulations treat kinetic neutrino transport in BNSMs and CCSNe separately from general relativistic hydrodynamic calculations. Consequently, the EOS tables used in hydrodynamics omit neutrino contributions.
Thus, adding trapped neutrinos in equilibrium with the rest of the matter results in loss of generality since it describes only systems in which neutrinos are in equilibrium (e.g. sometimes PNSs are described assuming neutrinos in equilibrium, see e.g. \cite{Logoteta:2022hvb}). In its current state, our table can be used both in simulations and as a starting point to incorporate neutrinos in or out of equilibrium.

This framework can also be generalized to include other continuous parameters controlling the local vs global conservation of charges besides electric charge (e.g., strangeness or isospin). For example, in heavy ion collision calculations, strangeness and isospin can be conserved fully globally or fully locally (e.g., in \cite{Greiner:1987tg,Lavagno:2022orw}, isospin is locally conserved, while the strangeness is globally conserved). In equilibrium, there is no physical motivation for a locally conserved strong charge, as no long-range force is associated with it. However, with local conservation, one can describe in a thermodynamical framework the scenario in which a conserved charge cannot be exchanged between two phases, namely when the reactions responsible for transferring this charge between the phases are suppressed. Extending our framework, one can study the intermediate cases in which reactions exchanging conserved charges, while allowed, are not fast enough to keep the two phases in chemical equilibrium with respect to that conserved charge. 

%

\begin{acknowledgments}
C.C. acknowledges support from the European Union's Horizon 2020 Research and Innovation Program under the Marie Sk\l{}odowska-Curie Grant Agreement No. 754496 (H2020-MSCA-COFUND-2016 FELLINI). 
T.Z. acknowledges support by the Network for Neutrinos, Nuclear Astrophysics and Symmetries (N3AS) through the National Science Foundation Physics Frontier Center, Grant No. PHY-2020275. 
S.H.'s work was supported by Startup Funds from the T.D. Lee Institute and Shanghai Jiao Tong University. 
M.P. acknowledges support by the Department of Energy, Award No. DE-FG02-93ER40756. 
The authors gratefully acknowledge the INT program ``Neutron Rich Matter on Heaven and Earth'' (INT-22r-2a) and the joint INT-N3AS workshop ``EOS Measurements with Next-Generation Gravitational-Wave Detectors'' (INT-24-89W) both held at the Institute for Nuclear Theory, University of Washington for hospitality and stimulating discussions. This research was supported in part by the INT's U.S. Department of Energy grant No. DE-FG02-00ER41132. 
\end{acknowledgments}

\appendix
\section{Surface tension and Debye screening length 
\label{apd:screen}}
In this appendix, we estimate the validity of LCN and GCN from the screening of electrons. The role of screening effects depends critically on the relative sizes of the quark droplets and the Debye screening length of electrons. When droplet sizes are small compared to the Debye screening length, the electron density remains largely uniform. In this regime, screening effects are negligible and GCN condition holds. In the regime where droplet sizes are much larger than the Debye screening length, the strong screening narrows the transition region, leaving the majority of the matter charge neutral in both phases. 

The size of nuclear droplets in deconfined quark matter can be estimated by balancing the surface energy and Coulomb energy, while neglecting the complexities introduced by the underlying quark-hadron interactions. The total energy of a droplet is expressed as: 
\begin{eqnarray}
    E_{tot}&=&4\pi R^2 \sigma+\frac{3}{5}\frac{Q^2}{4\pi R}
\end{eqnarray}
where $\sigma$ represents surface tension, $Q$ and $R$ denote the electric charge and radius of the droplet, respectively. The droplet charge is given by $Q=\Delta n_e V_Q\propto R^3$, where $\Delta n_e$ denotes the charge density difference between the inside and outside of the droplet, and $V_Q=\frac{4}{3}\pi R^3$ is the droplet's volume.  Minimizing the total energy per charge with fixed charge density 
, $d (E_{tot}/Q)/d R =0$ determines the equilibrium radius \cite{Ravenhall:1983uh,shapiro2024black,Glendenning:2012com}:
\begin{eqnarray}
R&=&\left(\frac{3Q^2}{40\pi^2\sigma}\right)^{1/3}
\end{eqnarray} 
Assuming deconfined quark matter is approximately charge neutral without leptons and its volume fraction $1-\chi$ is small, the charge $Q$ of the droplet can be estimated as: $Q=e Y_e \nb V_Q$. Substituting this expression for $Q$, the droplet radius becomes:
\begin{eqnarray}
    R &=& \left( \frac{15\sigma}{ 8\pi \alpha\left( Y_\text{e} \nb \right)^2} \right)^{1/3}\\
&=& 12.8\,[\textrm{fm}] \left (\frac{\sigma}{50\,[\textrm{MeV fm}^{-2}]}\right)^{1/3}
\left (\frac{Y_\text{e} \nb}{0.1\,[\textrm{fm}^{-3}]}\right)^{-2/3}
\end{eqnarray}
where $\alpha=e^2/{4\pi}$ is the fine-structure constant in Heaviside-Lorentz units. 

For comparison the electron Debye screening length, $\lambda_D$, is given by~\cite{Haensel:2007yy,Voskresensky:2002hu,{Heiselberg:1992dx}},
\begin{eqnarray}
    \lambda_D^{-1} &=& e\left(\frac{\partial n_\text{e}}{\partial \mu_\text{e}}\right)^{1/2}=\frac{e~k_{F,\text{e}}}{\pi}
\end{eqnarray}
where $k_{F,e}$ is the electron Fermi momentum (assuming a massless electron, $m_e = 0$). As before, taking the volume fraction of the nucleonic matter $\chi\approx 1$, we can write
\begin{eqnarray}
\lambda_D&=&\frac{\sqrt{\pi/4\alpha}}{(3\pi^2 Y_\text{e} \nb)^{\frac{1}{3}}}\\
&=& 7.2[\textrm{fm}] \left (\frac{Y_\text{e} \nb}{0.1[\textrm{fm}^{-3}]}\right)^{-1/3}
\end{eqnarray}

In the critical case where $R\approx\lambda_D$, the surface tension is: 
\begin{eqnarray}
    \sigma_{c} &=& \frac{ Y_\text{e} \nb \sqrt{\pi} }{45 \sqrt{\alpha}}\\
    &=& 9.1[\textrm{MeV/fm}^2] \left (\frac{Y_\text{e} \nb}{0.1[\textrm{fm}^{-3}]}\right)
\end{eqnarray}

For surface tensions $\sigma\gg\sigma_c$, large droplets form which follow the LCN condition. Conversely, for $\sigma\lesssim \sigma_c$, a nearly uniform, globally neutral sea of leptons and strongly-interacting matter is energetically favorable. However, $\sigma_c$ does not serve as a strict threshold distinguishing local from global charge neutrality. Instead, its strong dependence on the ratio $\sigma_c\propto \left({R}/{\lambda_D}\right)^3$  suggests that a broad range of surface tensions $\sigma$ lead to to various pasta phases, ranging from 1D slabs to 3D bubbles, which remain stable at different densities \cite{Maruyama:2007ey,Yasutake:2014oxa,Wu:2018zoe,Maslov:2018ghi,Ju:2021hoy}.

\section{JEL derivatives}\label{apd:JEL}
In this appendix we collect explicit derivatives of the JEL functions with respect to the variables $\nb$, $f$, and $g$ that are necessary for the numerical evaluation of the specific heats and the adiabatic sound speed.

First, we note that
\beq
\frac{df}{d\psi} &=& \left(\frac{d\psi}{df}\right)^{-1}
                  =\frac{f}{(1+f/a)^{1/2}}   \\
\left.\frac{\p g}{\p f}\right|_t &=& \frac{t}{2(1+f)^{1/2}}=\frac{t^2}{2g} \\
\left.\frac{\p g}{\p t}\right|_f &=& (1+f)^{1/2} =\frac{g}{t}   \\
\left.\frac{\p}{\p \psi}\right|_t &=& 
    \frac{df}{d\psi}\left(\left.\frac{\p}{\p f}\right|_g
    +\left.\frac{\p g}{\p f}\right|_t \left.\frac{\p}{\p g}\right|_f \right) \\
\left.\frac{\p}{\p t}\right|_{\psi} &=& 
    \left.\frac{\p g}{\p t}\right|_f \left.\frac{\p}{\p g}\right|_f   ~. 
\eeq
\subsection{Partial derivatives of $y$ with respect to $\nb$, $f$, and $g$}

\be
\left.\frac{\p y}{\p \nb}\right|_{f,g} = -\frac{y}{\nb}
\ee
\beq
\left.\frac{\partial y}{\partial f}\right|_{\nb,g} &=&  
\frac{\gamma}{\pi^2}\frac{m^3}{\nb}
\frac{fg^{3/2}(1+g)^{3/2}}{(1+f)^{M+1/2}(1+g)^N(1+f/a)^{1/2}} 
\nonumber \\
&\times& \sum_{m=0}^M\sum_{n=0}^Np_{mn}f^mg^n\left\{\frac{}{} \right.  
\nonumber \\
&&\left[1+m+\left(\frac{1}{4}+\frac{n}{2}-M\right)\frac{f}{1+f} \right. 
\nonumber \\
&&+\left.\left(\frac{3}{4}-\frac{N}{2}\right)\frac{fg}{(1+f)(1+g)}\right] \nonumber  \\
&&\times \left[\frac{1+m}{f}-\frac{1}{2a(1+f/a)}-\frac{M+1/2}{1+f}\right]     \nonumber  \\
&&+\left. \frac{1}{(1+f)^2}
\left[\frac{1}{4}+\frac{n}{2}-M+\left(\frac{3}{4}-\frac{N}{2}\right)
\frac{g}{1+g}\right]\right\}     \nonumber \\
\eeq
\beq
\left.\frac{\partial y}{\partial g}\right|_{\nb,f} &=&  
\frac{\gamma  }{\pi^2}\frac{m^3}{\nb}
\frac{fg^{3/2}(1+g)^{3/2}}{(1+f)^{M+1/2}(1+g)^N(1+f/a)^{1/2}} 
\nonumber \\
&\times& \sum_{m=0}^M\sum_{n=0}^Np_{mn}f^mg^n\left\{\frac{}{} \right.  
\nonumber \\
&&\left[1+m+\left(\frac{1}{4}+\frac{n}{2}-M\right)\frac{f}{1+f} \right. 
\nonumber \\
&&+\left.\left(\frac{3}{4}-\frac{N}{2}\right)\frac{fg}{(1+f)(1+g)}\right]
\nonumber  \\
&&\times \left[\left(\frac{3}{2}+n\right)\frac{1}{g}
+\left(\frac{3}{2}-N\right)\frac{1}{1+g}\right]  
\nonumber \\  
&&+\left.\frac{f}{(1+f)(1+g)^2}\left(\frac{3}{4}-\frac{N}{2}\right)\right\}   
\eeq
\subsection{Partial derivatives of $I_{\ep}^-$ with respect to $f$ and $g$}
\beq
\left.\frac{\p I_{\ep}^-}{\p f}\right|_g &=&
\left[\frac{1}{f}-\frac{M+1}{1+f}\right]I_{\ep}^-   \nonumber \\
&+&\frac{\gamma m^4}{\pi^2}\frac{fg^{5/2}(1+g)^{3/2}}{(1+f)^{M+1}(1+g)^N}
\sum_{m=0}^M\sum_{n=0}^Np_{mn}f^mg^n  \nonumber \\
&\times& \frac{m}{f}\left\{
\left[\frac{3}{2}+n+\left(\frac{3}{2}-N\right)\frac{g}{1+g}\right]\right\} 
\eeq
\beq
\left.\frac{\p I_{\ep}^-}{\p g}\right|_f &=&
\left[\frac{5}{2g}+\frac{3/2-N}{1+g}\right]I_{\ep}^- \nonumber \\
&+&\frac{\gamma m^4}{\pi^2}\frac{fg^{5/2}(1+g)^{3/2}}{(1+f)^{M+1}(1+g)^N}
\sum_{m=0}^M\sum_{n=0}^Np_{mn}f^m g^n \nonumber \\
&\times&\left\{\frac{n}{g}
\left[\frac{3}{2}+n+\left(\frac{3}{2}-N\right)\frac{g}{1+g}\right]\right. \nonumber \\
&+&\left.\left(\frac{3}{2}-N\right)\frac{1}{(1+g)^2}\right\} 
\eeq      
\subsection{Partial derivatives of $I_P$ with respect to $f$ and $g$}
\beq
\left.\frac{\p I_P}{\p f}\right|_g &=&
\left[\frac{1}{f}-\frac{M+1}{1+f}\right]I_P   \nonumber \\
&+&\frac{\gamma m^4}{\pi^2}\frac{fg^{5/2}(1+g)^{3/2}}{(1+f)^{M+1}(1+g)^N}
\sum_{m=0}^M\sum_{n=0}^Np_{mn}f^mg^n \frac{m}{f}  \nonumber \\
\eeq
\beq
\left.\frac{\p I_P}{\p g}\right|_f &=&
\left[\frac{5}{2g}+\frac{3/2-N}{1+g}\right]I_P \nonumber \\
&+&\frac{\gamma m^4}{\pi^2}\frac{fg^{5/2}(1+g)^{3/2}}{(1+f)^{M+1}(1+g)^N}
\sum_{m=0}^M\sum_{n=0}^Np_{mn}f^m g^n \frac{n}{g} \nonumber \\
\eeq      

\bibliography{bibliography}

\end{document}